\documentclass[numsec,webpdf,modern,medium,namedate]{oup-authoring-template}
\onecolumn 

\usepackage{hyperref}

\RequirePackage{amsthm,amsmath,amsfonts,amssymb}

\makeatletter
\def\ps@opening{%
  \def\@oddfoot{\sffamily\normalsize\hfill\thepage}%
  \def\@evenfoot{\normalfont\normalsize\thepage\hfill}%
  \let\@evenhead\relax
  \let\@oddhead\relax
}
\makeatother
\begin{document}
\journaltitle{Journal Title Here}
\DOI{DOI HERE}
\copyrightyear{2022}
\pubyear{2019}
\access{Advance Access Publication Date: Day Month Year}
\appnotes{Paper}
\firstpage{1}


\title[Separable models for dynamic signed networks]{Separable models for dynamic signed networks}
\author[$\ast$]{Alberto Caimo\ORCID{0000-0001-8956-7166}}
\author{Isabella Gollini\ORCID{0000-0002-7738-5688}}
\authormark{Caimo and Gollini}
\address{\orgdiv{School of Mathematics and Statistics}, \orgname{University College Dublin}, \orgaddress{\street{Belfield}, \postcode{D04 V1W8}, \country{Ireland}}}
\corresp[$\ast$]{Corresponding author. \href{email:alberto.caimo1@ucd.ie}{alberto.caimo1@ucd.ie}}

\abstract{Signed networks capture the polarity of relationships between nodes, providing valuable insights into complex systems where both supportive and antagonistic interactions play a critical role in shaping the network dynamics. 
We propose a separable temporal generative framework based on multi-layer exponential random graph models, characterised by the assumption of conditional independence between the sign and interaction effects. This structure preserves the flexibly and explanatory power inherent in the binary network specification while adhering to consistent balance theory assumptions. Using a fully probabilistic Bayesian paradigm, we infer the doubly intractable posterior distribution of model parameters via an adaptive Metropolis-Hastings approximate exchange algorithm. We illustrate the interpretability of our model by analysing signed relations among U.S. Senators during Ronald Reagan's second term (1985–1989). Specifically, we aim to understand whether these relations are consistent and balanced or reflect patterns of supportive or antagonistic alliances.}
\maketitle

\section{Introduction}

Network data characterise the connectivity structures of complex systems, which often involve a delicate balance between two types of interactions. In biological systems, for example, these interactions may be reflected in co-expression similarities and dissimilarities between genes \citep{masetal09} that are explained by the sign of the correlation of their expression profiles, while in social or political networks, they manifest as alliances and rivalries between individuals \citep{fontan2021}. 

Signed networks, which consist of positive and negative edges, provide a powerful tool for capturing the nuanced relationships within these systems. In fact, by using signed networks, we can identify patterns and trends that reveal deeper insights into the dynamics of these complex systems, which may not be immediately visible through other analytical methods.

A central question in the analysis of signed networks is how well positive and negative edges align with Structural Balance Theory (SBT) \citep{car:har56}. SBT asserts that positive relations should exhibit transitivity (two positive paths should close with a positive edge) while negative relations follow anti-transitivity (two negative paths should close with a positive edge). Conversely, a positive two-path should not be closed by a negative relation, and a negative two-path should not be closed by another negative relation, as these patterns produce unbalanced triads.

To explore these dynamics, our approach utilises the exponential random graph modelling (ERGM) framework \citep{lus:kos:rob13}, which is particularly suited to modelling the complex interplay between positive and negative edge configurations. ERGMs allow us to incorporate terms that capture relational tendencies, such as the clustering of positive edges or the formation of unexpected negative edges between individuals with similar nodal covariate information. By accounting for these relational patterns, we can accurately assess the statistical significance of balanced configurations and their impact on the overall structure of the network \citep{fri:meh:thu:kau22}.

In this paper, we examine undirected signed network panel data, which consists of cross-sectional snapshots of network relationships observed at regular intervals. In particular, we focus on the analysis of political relationships among U.S. Senators, measured across successive U.S. Congresses \citep{zac14, zac20}. Each snapshot captures the state of signed interactions within the Senate during each legislative year, offering insight into the evolving dynamics of political alliances and rivalries over time.
We extend the temporal exponential random graph model (TERGM) framework \citep{rob:pat01, han:fu:xin10} to better accommodate the dynamics of signed interactions in longitudinal network data. Specifically, we incorporate a separable specification \citep{kri:han14} that distinguishes between the formation and persistence of edges, enabling a more detailed understanding of temporal dependence. 
Building on the decomposition introduced by \citet{ler16}, we further enhance the TERGM by modelling the signed network as the outcome of two interrelated processes: an interaction process and a conditional sign process. This extension allows for a more efficient and interpretable parameterisation that captures not only how relationships form and persist over time, but also how structural balance, through both balanced and imbalanced signed configurations, shapes the evolution of network interactions.

This perspective aligns with recent work showing that micro-level mechanisms can accumulate to produce pronounced patterns at the macro level \citep{dux24}. Even modest biases in the formation or sign of individual edges can have substantial effects on overall network structure, influencing cohesion, clustering, and balance \citep{amaetal18}. Incorporating endogenous  micro-level effects into our separable framework ensures that both the local dynamics and their consequences for global network patterns are explicitly represented.

The paper is structured as follows. In Sections~\ref{sec:ergms} and~\ref{sec:tergms}, we provide an overview of ERGMs and temporal ERGMs respectively. Section~\ref{signed_nets} offers a comprehensive definition of signed networks and presents an overview of SBT. In Sections~\ref{sec:multi_ergms} and~\ref{sec:sepmod}, we introduce a multi-layer interpretation of the signed process, distinguishing between the interaction process and the conditional sign process. We describe how these two components are incorporated into the ERGM generative framework to capture the dynamics of signed relationships. Section~\ref{sec:2lSTERGMs} combines temporal separability with layer separability to define 2-layer separable TERGMs. Here, we define a Markov process based on binary ERGM terms, which allows for a principled handling of Structural Balance Theory configurations. In Section~\ref{Bayesian}, we outline a Bayesian inferential approach, which quantifies the uncertainty in the posterior model parameter estimates by using an adaptive approximate exchange algorithm \citep{haa01, cai:fri11} to facilitate sampling from the doubly-intractable posterior distribution. 
In Section \ref{sim}, we present a comparative simulation study to demonstrate parameter recoverability and the robustness of the multi-layer conditional approach, and to highlight the differences between this method, the TERGM approach of \citet{fri:meh:thu:kau22}, and its separable temporal variant.
Finally, Section~\ref{app} demonstrates the utility of our modelling approach by applying it to the analysis of political relationships among U.S. Senators. We assess model fit using posterior predictive checks to evaluate the model  performance. Section~\ref{discussion} concludes with a discussion on potential extensions and future directions.

\section{Exponential random graph models}\label{sec:ergms}

The relational structure of a network graph is described by a random adjacency matrix $\mathbf{y}$ of a graph on $n$ nodes (actors) and a set of edge variables such that $y_{ij} = 1,$ if node $i$ is connected to node $j$ ($i \sim j$); $y_{ij} = 0,$ if node $i$ is not connected to node $j$ ($i \not\sim j$).

Exponential random graph models (ERGMs) \citep{hol:lei81,str:ike90} are a particular class of discrete linear exponential families with probability mass function:
\begin{equation}\label{eq:ergm}
p(\mathbf{y} \mid \boldsymbol{\varphi}) = \exp\{\boldsymbol{\varphi}^\top \mathbf{s}(\mathbf{y}) - \kappa(\boldsymbol{\varphi})\},
\end{equation}
where $\mathbf{s}(\mathbf{y}) \in \mathbb{R}^Q_{>0}$ is a known vector of $Q$ sufficient statistics \citep{bes74}, $\boldsymbol{\varphi} \in \mathbb{R}^Q$ is the associated parameter vector, and $\kappa(\boldsymbol{\varphi})$ a normalising constant which is computationally intractable to evaluate for all but trivially small graphs. The dependence hypothesis at the basis of these models is that edges form small structures called configurations. There is a wide range of possible network configurations \citep{rob:pat:kal:lus07} which give flexibility to adapt to different contexts. A positive (or negative) value of the parameter $\varphi$ indicates a tendency for the configuration represented by $s(\mathbf{y})$ to appear more (or less) frequently in the data than would be expected under an Erdős–Rényi random graph model with an edge probability of 0.5.

The ERGM likelihood defined in Equation~\eqref{eq:ergm} can be generalised by allowing the parameter to vary non-linearly within the exponential family, resulting in a curved exponential family \citep{hun07}. Monte Carlo methods \citep{hun:han06} can be used to estimate the decay parameter of the geometrically weighted network statistics introduced by \cite{sni:pat:rob:han06} and currently used to alleviate ERGM degeneracy issues \citep{han03}. 

\section{Temporal exponential random graph models}\label{sec:tergms}

Temporal exponential random graph models (TERGMs) \citep{rob:pat01, han:fu:xin10} describe the joint distribution of a dynamic network sequence by assuming a Markov process on the network from one time step to the next:
$$p(\mathbf{y}_{1},\dots, \mathbf{y}_{T} \mid \mathbf{y}_{0}, \boldsymbol{\varphi}) = \prod_{t = 1}^T p\left(\mathbf{y}_t \mid \mathbf{y}_{t-1}, \boldsymbol{\varphi}\right) = \prod_{t = 1}^T \exp \{\boldsymbol{\varphi}^{\top} \mathbf{s}\left(\mathbf{y}_t; \mathbf{y}_{t-1}\right) - \kappa(\boldsymbol{\varphi}; \mathbf{y}_{t-1})\},$$
where $\boldsymbol{\varphi}$ parametrises the influence of sufficient network statistics $\mathbf{s}\left(\mathbf{y}_t; \mathbf{y}_{t-1}\right)$ on the likelihood and $\kappa(\boldsymbol{\varphi}, \mathbf{y}_{t-1})$ is a normalising constant. 
The process can be extended to incorporate higher-order Markov dependence by assuming that the network $\mathbf{y}_t$ depends on $K \in \{0, 1, \cdots, T - 1\}$ previously observed networks. This is achieved by including lagged networks in the network statistics $\mathbf{s}(\cdot)$ with $K$ selected  appropriately \citep{lei;cra:des18}.

The separable temporal parametrisation (STERGM) proposed by \cite{kri:han14} provides a convenient control over incidence and duration of edges and separate interpretation between consecutive network observations.
We define the formation network $\mathbf{y}^\mathcal{F} = \mathbf{y}_{t-1} \cup \mathbf{y}_{t}$;
and the persistence network $\mathbf{y}^\mathcal{P} = \mathbf{y}_{t-1} \cap \mathbf{y}_{t}$; and assume $\mathbf{y}^\mathcal{F}$ and $\mathbf{y}^\mathcal{P}$ are conditionally independent given $\mathbf{y}_{t-1}$ so that:
\begin{equation}\label{eq:stergm}
p\left(\mathbf{y}_t \mid \mathbf{y}_{t - 1}, \boldsymbol{\varphi}^\mathcal{F},\boldsymbol{\varphi}^\mathcal{P}\right) = p\left(\mathbf{y}^\mathcal{F} \mid \mathbf{y}_{t - 1}, \boldsymbol{\varphi}^\mathcal{F}\right)\times p\left(\mathbf{y}^\mathcal{P} \mid \mathbf{y}_{t - 1}, \boldsymbol{\varphi}^\mathcal{P}\right),
\end{equation}
where $\boldsymbol{\varphi}^\mathcal{F}$ and $\boldsymbol{\varphi}^\mathcal{P}$ describe respectively the edge formation and persistence process from $t-1$ to $t.$ Then, given $\mathbf{y}^\mathcal{F}, \mathbf{y}^\mathcal{P}$ and $\mathbf{y}_{t-1},$ we have $\mathbf{y}_t=\mathbf{y}^\mathcal{P} \cup\left(\mathbf{y}^\mathcal{F}\; \backslash\; \mathbf{y}_{t-1}\right).$

The STERGM framework addresses a key limitation of standard TERGMs by separating the modelling of edge formation and persistence. This separation allows for more interpretable modelling of dynamic networks, especially when we assume that edge formation and dissolution are governed by different mechanisms. For example, in a friendship network, a new friendship (formation) may be influenced by proximity while the maintenance of an existing friendship (persistence) may supported by the presence of mutual friends.

\section{Signed network structures}\label{signed_nets}

A signed network consists of a set of $n$ nodes and signed edges between pairs of them. Its relational structure is described by a random adjacency matrix with elements of the set of edge variables $y_{ij} \in \{+1, -1, 0\}$ respectively indicating a positive, negative or no relation between node $i$ and node $j.$ As for binary networks, signed networks may be directed or undirected depending on the nature of the relationships between the nodes. In this paper, we focus on signed undirected networks, although the methodologies introduced can be easily extended to the directed case.

\begin{figure}[htp]
\centering
\includegraphics[scale=0.25]{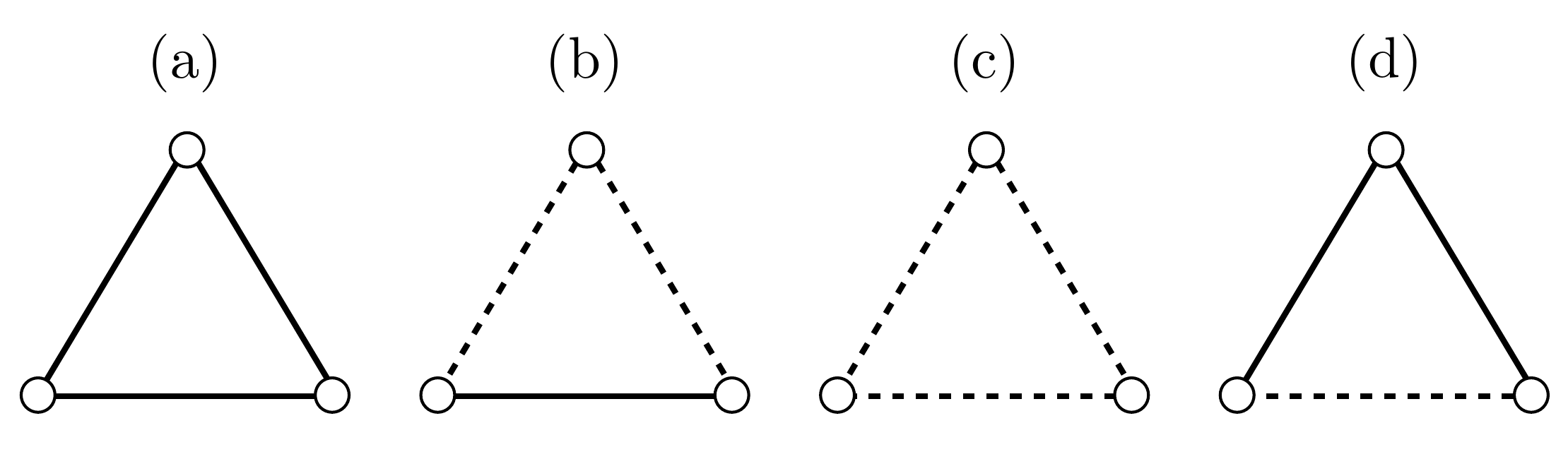}
\caption{Triadic configurations of a signed network. Solid lines represent positive interactions,  dashed lines represent negative interactions.}
\label{fig:SBT}
\end{figure}

Structural balance theory (SBT) examines how positive and negative relationships between actors evolve and ultimately stabilise within a network. It focuses on the dynamics and eventual equilibrium of these relationships in a signed network \citep{car:har56}. SBT is commonly illustrated using the familiar concepts of friendship and enmity. These terms provide a clear and intuitive understanding of the fundamental principles of the theory. Balanced triadic configurations include: (1) a friend of a friend is a friend, and (2) an enemy of a friend is an enemy. In contrast, unbalanced configurations are: (3) a friend of an enemy is an enemy, and (4) an enemy of an enemy is a friend. Configuration (1) corresponds to positive triangles (Figure~\ref{fig:SBT}~(a)), while configuration (2) corresponds to triangles with two negative edges (Figure~\ref{fig:SBT}~(b)). These two patterns are part of the so-called strong balance structure \citep{har53}. \cite{dav67} later introduced the weak balance structure, under which configurations comprising triangles with all negative edges (Figure~\ref{fig:SBT}~(c)) are also regarded as balanced. Conversely, triadic configurations with exactly two positive edges and one negative edge (Figure~\ref{fig:SBT}~(d)) are considered unbalanced.

The underlying motivation for structural balance is dynamic and based on how unbalanced triangles should resolve to balanced ones. This situation has led naturally to a search for a full dynamic theory of structural balance. Yet finding systems that reliably guide networks to balance has proved to be a challenge in itself \citep{mar:kle:kle:str11}.

\section{Multi-layer interpretation}\label{sec:multi_ergms}

Multi-layer ERGMs \citep{kri:koe:mar20} extend the traditional ERGM framework to model networks with multiple types of interactions between nodes, with each layer representing a distinct kind of relationship. 
Each layer is modelled using its own set of network statistics, capturing the structural tendencies specific to that type of interaction, while allowing for dependencies across layers.

For example, suppose $\mathbf{Y}_1$ represents a binary friendship network and $\mathbf{Y}_2$ represents a binary collaboration network among the same set of individuals. In this setting, any dyad can simultaneously exhibit both types of relations (e.g., two individuals may be friends, collaborators, or both). A multi-layer ERGM can include statistics such as transitivity or homophily within each layer (e.g., friends-of-friends in $\mathbf{Y}_1$, shared collaborators in $\mathbf{Y}_2$), while also incorporating cross-layer effects, such as whether an edge in $\mathbf{Y}_1$ increases the likelihood of an edge in $\mathbf{Y}_2$.

In the context of signed networks, where a pair of nodes cannot simultaneously share both a positive and a negative edge, the signed edge process can be decomposed as $\mathbf{Y} = \mathbf{X} \mathbf{Z},$ where $\mathbf{X},$ with dyadic elements $X_{ij} = 1$ if an edge exists between nodes $i$ and $j$ and $X_{ij} = 0$ otherwise, is a binary adjacency matrix encoding edge presence, and $\mathbf{Z},$ with dyadic elements $Z_{ij} = +1$ for positive edges and $Z_{ij} = -1$ for negative edges, encodes edge signs. This formulation ensures that each dyad has at most one edge with a single sign. 
Clearly, this can be viewed as a special case of a multi-layer network, where the sign information is meaningful only in the presence of an actual interaction. In other words, the sign layer is effectively constrained by the interaction layer. This dependency will be further discussed in the following section.

However, one could also argue that signed relations may exist independently of observed interactions, i.e., the sign of a relationship could be latent, with the interaction merely serving to reveal or activate a pre-existing evaluative stance. In this view, the act of interaction exposes underlying sentiments such as trust, hostility, or alliance that already shape the relational landscape. This interpretation aligns, for example, with sociological theories of latent affective edges \citep{heider1946} and is particularly relevant in contexts where relationships are long-standing or embedded in broader social structures. In such settings, the absence of interaction does not necessarily imply neutrality or indifference, but rather a dormant state of the relationship.

We assume that the observed signed network follows a 2-layer ERGM with the probability mass function:
\begin{equation}\label{eq:ERGM_2}
p\left(\mathbf{y} \mid \boldsymbol{\vartheta}\right) =
\exp \left\{ 
\boldsymbol{\vartheta}^{\top} \mathbf{s}(\mathbf{x}, \mathbf{z}) - \kappa(\boldsymbol{\vartheta}, \mathbf{x}, \mathbf{z})\right\},
\end{equation}
where  
$\mathbf{s}(\mathbf{x}, \mathbf{z})$ incorporates network statistics that model the two layers jointly (see Figure~\ref{fig:model_graphs} (a)). The modelling approach for signed networks proposed by \cite{fri:meh:thu:kau22} is equivalent to a 2-layer ERGM which represents an ERGM with a uniform categorical reference measure where each dyad can take one of three possible states: positive, negative, or absent. The conditional probability that a dyad $(i,j)$ is positive, given the rest of the network $\mathbf{y}_{-(ij)}$ and the model parameter $\boldsymbol{\vartheta}$, is
$$
\Pr\left(Y_{ij}= +1 \mid \mathbf{y}_{-(ij)}, \boldsymbol{\vartheta} \right)=\frac{\exp\left\{\boldsymbol{\vartheta}^{\top}\mathbf{s}\left(\mathbf{y}_{ij}^{+1}\right)\right\}}{\sum_{y\in\{+1,-1,0\}}\exp\left\{\boldsymbol{\vartheta}^{\top}\mathbf{s}\left(\mathbf{y}_{ij}^{y}\right)\right\}},
$$
where $\mathbf{s}(\mathbf{y}_{ij}^{y})$ is the vector of network statistics evaluated when dyad $(i,j)$ is in state $y$.

\section{Separable modelling}\label{sec:sepmod}

\cite{ler16} highlights the distinction between marginal and conditional tests of structural balance, showing that other network effects influencing the likelihood of interaction can mask the presence of balance. In such cases, structural balance may appear absent even when nodes actually prefer balanced triangles over unbalanced ones. To address this, a proper test of balance theory requires modelling the conditional probability of an edge being positive or negative, given that the edge exists. More recently, \citet{ler:lom20} have shown that, when conditioning on active users within a large relational event framework, balance theory provides a compelling explanation for the network structure of teams engaged in contentious tasks.

\begin{figure}[htp]
\centering
\includegraphics[scale=0.63]{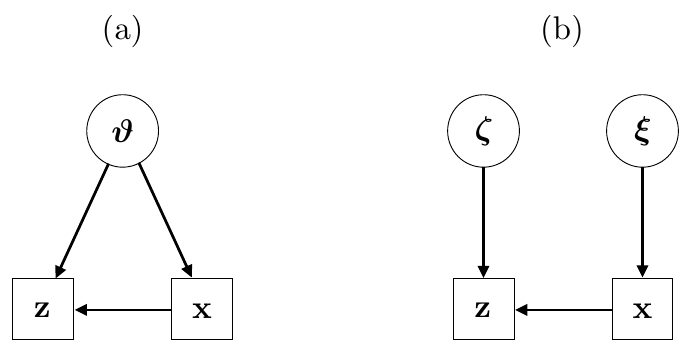}
\caption{Graphical representation of (a) 2-layer ERGMs defined in Equation~\eqref{eq:ERGM_2} and (b) separable 2-layer ERGMs defined in Equation~\eqref{eq:sepERGM_2}. Squares represent observed variables (network layers); circles represent parameters and arrows represent directed dependencies. In panel (a), both $\mathbf{x}$ and $\mathbf{z}$ are influenced by a common parameter $\boldsymbol{\vartheta}.$ In panel (b), 
$\mathbf{x}$ and $\mathbf{z}$ each have their own parameters, denoted by $\boldsymbol{\xi}$ and $\boldsymbol{\zeta}$, respectively. These parameters are distinct, meaning the two processes are conditionally independent given $\mathbf{x}.$}
\label{fig:model_graphs}
\end{figure}

In line with this insight, our model explicitly separates the modelling of the interaction process ($\mathbf{X}$) from the sign process ($\mathbf{Z}$). Unlike the joint model described in the previous section, this separable approach allows the edge structure and edge signs to be modelled conditionally independently, while still accounting for their dependence through the conditional formulation.
As shown by \cite{ler16} through analyses of multiple benchmark datasets commonly used in the literature, this separation offers both practical and interpretative advantages. It simplifies model estimation, and it reflects situations where the drivers of edge formation, such as affiliations, or spatial proximity, are distinct from those influencing edge polarity, such as trust, cooperation, or past conflict. For example, in a social network, whether an edge exists may be determined by organisational or logistical constraints, whereas the sign of that edge depends on the nature of the relationship.
Formally, the marginal probability of the network $\mathbf{Y}$ can be decomposed as the product of the probability of interaction $\mathbf{X}$ and the conditional probability of the sign given interaction $\mathbf{Z} \mid \mathbf{X}$:
$$
\Pr(\mathbf{Y} = \mathbf{y}) = \Pr(\mathbf{Z} = \mathbf{z} \mid \mathbf{X} = \mathbf{x}) \; \Pr(\mathbf{X} = \mathbf{x}).
$$
The fact that $\mathbf{Z}$ is conditional on $\mathbf{X}$ does not imply a temporal ordering. The conditionality reflects the structural dependence of edge signs on the presence of edges and does not indicate that $\mathbf{Z}$ evolves after $\mathbf{X}$.
This explicit separation clarifies how structural effects can be estimated for each process without conflating the mechanisms driving edge formation and edge sign. 
This leads to the following formulation of a separable ERGM:
\begin{equation}\label{eq:sepERGM_2}
\begin{aligned}[b]
p\left(\mathbf{y} \mid \boldsymbol{\zeta},\boldsymbol{\xi}\right) &= 
p\left(\mathbf{x} \mid \boldsymbol{\xi}\right)\times p\left(\mathbf{z} \mid \mathbf{x}, \boldsymbol{\zeta}\right) \\ 
&= \exp \left\{\boldsymbol{\xi}^{\top} \mathbf{s}\left(\mathbf{x} \right) - \kappa(\boldsymbol{\xi}) \right\} \times \exp \left\{\boldsymbol{\zeta}^{\top} \mathbf{s}\left(\mathbf{z}; \mathbf{x}\right) - \kappa(\boldsymbol{\zeta}, \mathbf{x}) \right\},
\end{aligned}
\end{equation}
where $\boldsymbol{\xi}$ governs the marginal interaction process, while $\boldsymbol{\zeta}$ governs the conditional sign process within the interaction structure. Figure~\ref{fig:model_graphs} (b) provides a graphical representation of the model, illustrating that the two parameters, $\boldsymbol{\xi}$ and $\boldsymbol{\zeta}$, are conditionally independent given the interaction process $\mathbf{x}$. Consequently, we define Model~\eqref{eq:sepERGM_2} as a \textsl{separable 2-layer ERGM.}

\section{Layered separable temporal models}\label{sec:2lSTERGMs}

The separable temporal assumption defined in Equation~\eqref{eq:stergm}, when combined with the 2-layer separability assumption specified in Equation~\eqref{eq:sepERGM_2}, results in a 2-layer STERGM, with the transition probability defined as:
\begin{equation}\label{eq:sep_stergm_2}
\begin{aligned}[b]
p\left(\mathbf{y}_t \mid \mathbf{y}_{t - 1}, \boldsymbol{\xi}^\mathcal{F}, \boldsymbol{\zeta}^\mathcal{F}, \boldsymbol{\xi}^\mathcal{P}, \boldsymbol{\zeta}^\mathcal{P}  \right)
= &\; p\left(\mathbf{y}^\mathcal{F}_t \mid  \mathbf{y}_{t - 1}, \boldsymbol{\xi}^\mathcal{F}, \boldsymbol{\zeta}^\mathcal{F} \right) \times p\left(\mathbf{y}^\mathcal{P}_t \mid \mathbf{y}_{t - 1}, \boldsymbol{\xi}^\mathcal{P},\boldsymbol{\zeta}^\mathcal{P}  \right)\\
= &\; p\left(\mathbf{z}_t^\mathcal{F} \mid \mathbf{x}_t^\mathcal{F}, \mathbf{y}_{t - 1}, \boldsymbol{\zeta}^\mathcal{F} \right) \times p\left(\mathbf{x}_t^\mathcal{F} \mid  \mathbf{y}_{t - 1}, \boldsymbol{\xi}^\mathcal{F} \right) 
       \times \\ 
  &\; p\left(\mathbf{z}_t^\mathcal{P} \mid \mathbf{x}_t^\mathcal{P}, \mathbf{y}_{t - 1}, \boldsymbol{\zeta}^\mathcal{P} \right) \times  p\left(\mathbf{x}_t^\mathcal{P} \mid  \mathbf{y}_{t - 1}, \boldsymbol{\xi}^\mathcal{P} \right)
       \\
= & \exp \left\{ {\boldsymbol{\zeta}^\mathcal{F}}^{\top} \mathbf{s}\left(\mathbf{z}^\mathcal{F}; \mathbf{x}^\mathcal{F}, \mathbf{y}_{t-1} \right) - \kappa({\boldsymbol{\zeta}^\mathcal{F}}; \mathbf{x}^\mathcal{F}, \mathbf{y}_{t - 1}) \right\} \times \\
& \exp \left\{ {\boldsymbol{\xi}^\mathcal{F}}^{\top} \mathbf{s}\left(\mathbf{x}^\mathcal{F}; \mathbf{y}_{t-1} \right) -\kappa({\boldsymbol{\xi}^\mathcal{F}}; \mathbf{y}_{t - 1}) \right\} \times \\
&\exp \left\{ {\boldsymbol{\zeta}^\mathcal{P}}^{\top} \mathbf{s}\left(\mathbf{z}^\mathcal{P}; \mathbf{x}^\mathcal{P}, \mathbf{y}_{t-1} \right) - \kappa({\boldsymbol{\zeta}^\mathcal{P}}; \mathbf{x}^\mathcal{P}, \mathbf{y}_{t - 1}) \right\} \times \\
&\exp \left\{ {\boldsymbol{\xi}^\mathcal{P}}^{\top} \mathbf{s}\left(\mathbf{x}^\mathcal{P}; \mathbf{y}_{t-1} \right) - \kappa({\boldsymbol{\xi}^\mathcal{P}}; \mathbf{y}_{t - 1})\right\}, 
\end{aligned}
\end{equation}
where $\boldsymbol{\xi}^\mathcal{F}$ and $\boldsymbol{\zeta}^\mathcal{F}$ describe the contribution of the interaction network effects and the edge sign network effects to the formation process, respectively; and where $\boldsymbol{\xi}^\mathcal{P}$ and $\boldsymbol{\zeta}^\mathcal{P}$ describe the contribution of the interaction network effects and the edge sign network effects to the persistence process, respectively. 

Model~\eqref{eq:sep_stergm_2} inherits the separability property within each time interval from Model~\eqref{eq:stergm} in a way that, in each interval $[t-1, t]$ the parameters  $\boldsymbol{\xi}$ and $\boldsymbol{\zeta}$ are conditionally independent given the current interaction structure $\mathbf{x}_t$ and the past information $\mathbf{y}_{t-1}.$ This means that while the formation and sign layers are modelled as conditionally independent within each time interval, Model~\eqref{eq:sep_stergm_2} does not necessarily imply full temporal independence across time. In fact, the structure of $\mathbf{x}_t$ may be influenced not only by $\mathbf{x}_{t-1}$, but also by $\mathbf{z}_{t-1}$, and similarly, $\mathbf{z}_t$ may depend not only on $\mathbf{z}_{t-1}$ but also on $\mathbf{x}_{t-1}$. In other words, the presence or absence of an interaction at time $t-1$ may explain not only the formation or dissolution of an edge at time $t$, but also changes in its sign; likewise, the sign structure at $t-1$ may influence not only whether sign persist or switch, but also whether new edges appear or existing ones disappear. This form of cross-layer temporal dependence is a key feature of the model, and reflects the fact that interaction and sign dynamics are often intertwined in practice. To achieve \textsl{full separability over time} between the two processes, we need to assume that at any time point $t,$ $\mathbf{x}^\mathcal{F}$ and $\mathbf{x}^\mathcal{P}$ (and therefore $\mathbf{x}$) do not depend on $\mathbf{z}_{t-1},$ i.e., at any time point $t,$ $\mathbf{s}\left(\mathbf{x}^\mathcal{F}; \mathbf{y}_{t-1} \right) = \mathbf{s}\left(\mathbf{x}^\mathcal{F}; \mathbf{x}_{t-1} \right)$ and $\mathbf{s}\left(\mathbf{x}^\mathcal{P}; \mathbf{y}_{t-1} \right) = \mathbf{s}\left(\mathbf{x}^\mathcal{P}; \mathbf{x}_{t-1} \right).$ 

\begin{figure}[htp]
\centering
\includegraphics[scale=0.75]{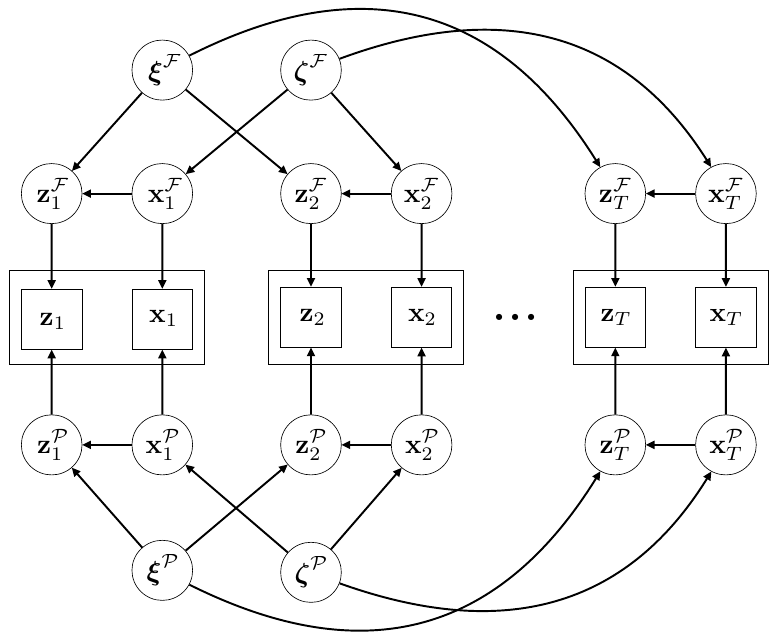}
\caption{Graphical representation of 2-layer STERGMs for signed network data conditional on an initial network state (not included).}
\label{fig:STERGMsep_2_graph}
\end{figure}

Figure~\ref{fig:STERGMsep_2_graph} provides a graphical representation of the 2-layer STERGM and illustrates that, given the interaction processes $\mathbf{x}_1$ and $\mathbf{x}_2$, the corresponding formation and persistence interaction processes, $\mathbf{x}_1^\mathcal{F}\), \(\mathbf{x}_2^\mathcal{F}\), \(\mathbf{x}_1^\mathcal{P}\), and \(\mathbf{x}_2^\mathcal{P}$, are fixed. This implies that \(\boldsymbol{\xi}^\mathcal{F}\) and \(\boldsymbol{\zeta}^\mathcal{F}\) are conditionally independent, as are \(\boldsymbol{\xi}^\mathcal{P}\) and $\boldsymbol{\zeta}^\mathcal{P}.$ However, the temporal dependence between \(\boldsymbol{\xi}^\mathcal{F}\) and \(\boldsymbol{\xi}^\mathcal{P}\), as well as between \(\boldsymbol{\zeta}^\mathcal{F}\) and \(\boldsymbol{\zeta}^\mathcal{P}\), still holds.

Table~\ref{tab:trans} presents the structure of all possible dyadic transitions in the signed network process over a given time interval, capturing how the relationships between node pairs evolve in terms of sign and connectivity. We can therefore interpret the conditional probability of observing a positive relation between nodes $i$ and $j$ at time $t$ in either the formation or persistence process by examining the corresponding log-odds expression:
\begin{equation}\label{eq:or}
\log \left( 
\frac{\Pr(Z^\mathcal{U}_{ij,t} = +1 \mid \mathbf{y}^\mathcal{U}_{-(ij),t}, X^\mathcal{U}_{ij,t} = 1, \mathbf{y}_{t-1})}
{\Pr(Z^\mathcal{U}_{ij,t} = -1 \mid \mathbf{y}^\mathcal{U}_{-(ij),t}, X^\mathcal{U}_{ij,t} = 1, \mathbf{y}_{t-1})} \right) = \boldsymbol{\zeta}^{\mathcal{U}^\top} \Delta_{ij} \mathbf{s}(\mathbf{z}_t^\mathcal{U}; \mathbf{x}_t^\mathcal{U}, \mathbf{y}_{t-1}), \text{ for } \mathcal{U} = \{\mathcal{F}, \mathcal{P}\}, 
\end{equation}
where $Z^\mathcal{U}_{ij,t}$ is the potential sign between $i$ and $j$ at time $t$ in either formation or dissolution process, given that the interaction at time $t$ exists in the corresponding process ($X^\mathcal{U}_{ij,t} = 1$); and the rest of the signed network in the respective process ($\mathbf{y}^\mathcal{U}_{-(ij),t}$). The right-hand side of the Equation~\eqref{eq:or} is the network change statistics that would result from toggling the sign of $Z_{ij,t}$, weighted by the parameter  $\boldsymbol{\zeta}^\mathcal{F}$ or $\boldsymbol{\zeta}^\mathcal{P}$, respectively. A higher log-odds means the formation or persistence of the sign of the interaction is more likely under the model, based on how well it matches the network configurations encouraged by the corresponding effects.

\begin{table}
\caption{(a) Possible transitions of a single signed edge variable $Y_{ij}$ from $t - 1$ to $t;$ (b) corresponding transitions according to the separable model defined in Equation~\eqref{eq:sep_stergm_2}. When no interaction is present $(X_{ij} = 0)$ the  associated sign $Z_{ij}$ is not identifiable, which in turn does not allow us to model possible changes in its value.}\label{tab:trans}
\setlength{\tabcolsep}{6pt}
\begin{tabular}{ccc|cccccccccc}
\hline \multicolumn{3}{c|}{(a)} & \multicolumn{10}{c}{(b)} \\
\hline $Y_{ij,t-1}$ & $\rightarrow$ & $Y_{ij,t}$   
                                & \multicolumn{2}{c}{$Y_{ij,t-1}$} &  
                                & \multicolumn{2}{c}{$Y_{ij}^\mathcal{F}$} 
                                & \multicolumn{2}{c}{$Y_{ij}^\mathcal{P}$} & 
                                & \multicolumn{2}{c}{$Y_{ij,t}$}       
                                \\
       \multicolumn{3}{c|}{} &
       $X_{ij,t-1}$ & $Z_{ij,t-1}$ & $\rightarrow$ & 
       $X_{ij}^\mathcal{F}$ & $Z_{ij}^\mathcal{F}$ &  
       $X_{ij}^\mathcal{P}$ & $Z_{ij}^\mathcal{P}$ & 
       $\rightarrow$  & $X_{ij,t}$ & $Z_{ij,t}$ \\
\hline $0$  & & $0$   & $0$ &      & & $0$ &      & $0$ &      & & $0$ &     \\
       $0$  & & $+1$  & $0$ &      & & $1$ & $+1$ & $0$ &      & & $1$ & $+1$ \\
       $0$  & & $-1$  & $0$ &      & & $1$ & $-1$ & $0$ &      & & $1$ & $-1$ \\
\hline $+1$ & & $0$   & $1$ & $+1$ & & $1$ & $+1$ & $0$ &      & & $0$ &     \\
       $+1$ & & $+1$  & $1$ & $+1$ & & $1$ & $+1$ & $1$ & $+1$ & & $1$ & $+1$ \\
       $+1$ & & $-1$  & $1$ & $+1$ & & $1$ & $+1$ & $1$ & $-1$ & & $1$ & $-1$ \\
\hline $-1$ & & $0$   & $1$ & $-1$ & & $1$ & $-1$ & $0$ &      & & $0$ &     \\
       $-1$ & & $+1$  & $1$ & $-1$ & & $1$ & $-1$ & $1$ & $+1$ & & $1$ & $+1$ \\
       $-1$ & & $-1$  & $1$ & $-1$ & & $1$ & $-1$ & $1$ & $-1$ & & $1$ & $-1$ \\
\hline
\end{tabular}
\end{table}

\subsection{Model terms}\label{sec:specs}

As with any ERGM-based modelling framework, the selection of network statistics is highly flexible and context-dependent. Commonly used statistics, such as edges, homophily, and transitivity, are typically included in the model to capture key structural features. STERGMs are specified by two potentially different sets of network statistics for formation and persistence.

The network statistics \(\mathbf{s}\left(\mathbf{z}^\mathcal{F}; \mathbf{x}^\mathcal{F}, \mathbf{y}_{t-1} \right)\) and \(\mathbf{s}\left(\mathbf{z}^\mathcal{P}; \mathbf{x}^\mathcal{P}, \mathbf{y}_{t-1} \right)\) 
include standard binary effects. These effects are based on the assumption that the layers \(\mathbf{z}^\mathcal{F}\) and \(\mathbf{z}^\mathcal{P}\) are endogenous, i.e., structures that are determined within the generative process itself, rather than being specified by external covariate information, while \(\mathbf{x}^\mathcal{F}\), \(\mathbf{x}^\mathcal{P}\), and \(\mathbf{y}_{t-1}\) are treated as exogenous. 
The network statistics \(\mathbf{s}\left(\mathbf{x}^\mathcal{F}; \mathbf{y}_{t-1} \right)\) and \(\mathbf{s}\left(\mathbf{x}^\mathcal{P}; \mathbf{y}_{t-1} \right)\) for the 2-layer STERGM~\eqref{eq:sep_stergm_2} assume that the layers \(\mathbf{x}^\mathcal{F}\) and \(\mathbf{x}^\mathcal{P}\) are endogenous, conditional on \(\mathbf{y}_{t-1}\).

\begin{figure}[htp]
\centering
\includegraphics[scale=0.3]{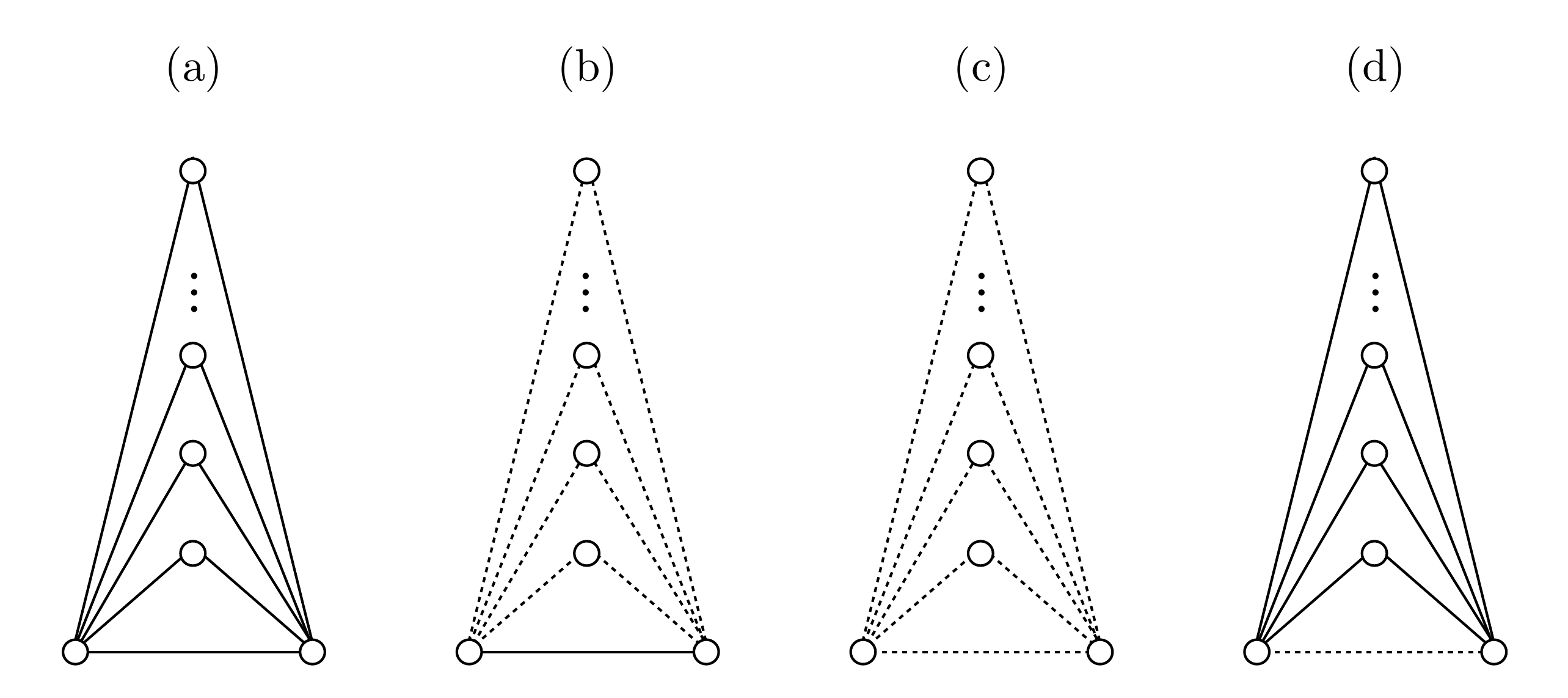}
\caption{Endogenous edgewise shared partners configurations: (a) \texttt{esf}$^{+}$: positive edgewise shared friends; (b) \texttt{ese}$^{+}$: positive edgewise shared enemies; (c) \texttt{ese}$^{-}$: negative edgewise shared enemies; (d) \texttt{esf}$^{-}$: negative edgewise shared friends.}
\label{SBT_stats}
\end{figure}

As clearly demonstrated by \cite{fri:meh:thu:kau22}, using lagged statistics that combine past and present edges can misrepresent the effects of SBT. For example, treating the presence of mutual friends or enemies at time $t-1$ as an exogenous covariate influencing the sign structure at time $t$ overlooks the fact that such covariates may not persist at time $t$, and may evolve in conjunction with other edge signs to form new balanced configurations. For this reason, in order to effectively model SBT effects, it is necessary to use endogenous network statistics that are implicitly dynamic, that is, they capture temporal dependencies without explicitly referencing past time points. It is important to emphasise that this reasoning also applies in our context, where SBT effects are estimated conditionally on the interaction structure. Indeed, in our case, assuming that such covariates remain constant would imply that both the interactions and their associated signs persist from time $t-1$ to time $t$, and are linked by an underlying structural dependency.

To test SBT effects, we adopt the geometrically-weighted statistics proposed by \cite{fri:meh:thu:kau22}. These statistics can be included as binary terms $\mathbf{s}\left(\mathbf{z}^\mathcal{F}; \mathbf{x}^\mathcal{F} \right)$ and $\mathbf{s}\left(\mathbf{z}^\mathcal{P}; \mathbf{x}^\mathcal{P} \right)$ in 2-layer STERGMs~\eqref{eq:sep_stergm_2} as the edgewise shared partner configurations shown in Figure~\ref{SBT_stats} are used conditionally on the presence of interactions, resulting in binary configurations within a constrained support. For example, the geometrically weighted distribution of negative edgewise shared friends, $\texttt{gwesf}^{-}(\mathbf{y}_t, \alpha)$, where $\alpha$ is the decay parameter, is equivalent to the binary geometrically weighted non-edgewise shared partners term, $\texttt{gwnsp}(\mathbf{z}_t; \mathbf{x}_t, \alpha)$, measured on $\mathbf{z}_t$ given $\mathbf{x}_t.$ 
In fact, conditional on $\mathbf{x}_t$, each dyad in $\mathbf{z}_t$ is binary, and therefore a negative edge that anchors the positive two-paths in the $\texttt{gwesf}^{-}$ configuration plays the same structural role as a non-edge anchoring the connected two-paths counted by the $\texttt{gwnsp}$ configuration in standard binary networks.

\section{Bayesian inference}\label{Bayesian}

We adopt a fully probabilistic Bayesian framework to estimate the 2-layer STERGM by defining a prior distribution over the model parameters \(\pi(\boldsymbol{\xi}^{\mathcal{F}}, \boldsymbol{\zeta}^\mathcal{F}, \boldsymbol{\xi}^{\mathcal{P}}, \boldsymbol{\zeta}^\mathcal{P})\), such that the posterior distribution is given by:
$$
\begin{aligned}[t]
\pi \left(\boldsymbol{\xi}^{\mathcal{F}}, \boldsymbol{\zeta}^\mathcal{F}, \boldsymbol{\zeta}^\mathcal{P}, \boldsymbol{\xi}^{\mathcal{P}} \mid \mathbf{y}_{0:T}\right) 
&\propto 
\pi(\boldsymbol{\xi}^{\mathcal{F}}, \boldsymbol{\zeta}^\mathcal{F}, \boldsymbol{\xi}^{\mathcal{P}}, \boldsymbol{\zeta}^\mathcal{P})
\prod_{t = 1}^T
p\left(\mathbf{y}_t \mid \mathbf{y}_{t - 1}, \boldsymbol{\xi}^{\mathcal{F}}, \boldsymbol{\zeta}^\mathcal{F}, \boldsymbol{\xi}^{\mathcal{P}}, \boldsymbol{\zeta}^\mathcal{P}  \right) \\
&= \pi(\boldsymbol{\xi}^\mathcal{F}, \boldsymbol{\zeta}^\mathcal{F}, \boldsymbol{\xi}^\mathcal{P}, \boldsymbol{\zeta}^\mathcal{P})
\prod_{t = 1}^T
p\left(\mathbf{z}_t \mid \mathbf{x}_t, \mathbf{y}_{t - 1}, \boldsymbol{\zeta}^\mathcal{F},\boldsymbol{\zeta}^\mathcal{P}  \right)
p\left(\mathbf{x}_t \mid  \mathbf{y}_{t - 1}, \boldsymbol{\xi}^\mathcal{F}, \boldsymbol{\xi}^\mathcal{P} \right).
\end{aligned}$$

By assuming independent priors $\pi(\boldsymbol{\xi}^\mathcal{F}, \boldsymbol{\zeta}^\mathcal{F}, \boldsymbol{\xi}^\mathcal{P}, \boldsymbol{\zeta}^\mathcal{P}) = \pi(\boldsymbol{\zeta}^\mathcal{F},\boldsymbol{\zeta}^\mathcal{P}) \times \pi(\boldsymbol{\xi}^\mathcal{F},\boldsymbol{\xi}^\mathcal{P})$ we preserve separability a posteriori:
$$
\pi \left(\boldsymbol{\xi}^\mathcal{F},\boldsymbol{\zeta}^\mathcal{F}, \boldsymbol{\xi}^\mathcal{P}, \boldsymbol{\zeta}^\mathcal{P} \mid \mathbf{y}_{0:T} \right)
=
\pi(\boldsymbol{\zeta}^\mathcal{F},\boldsymbol{\zeta}^\mathcal{P} \mid \mathbf{z}_{0:T}, \mathbf{x}_{0:T}) \times
\pi(\boldsymbol{\xi}^\mathcal{F},\boldsymbol{\xi}^\mathcal{P} \mid \mathbf{x}_{0:T}).
$$

The two posterior terms are doubly intractable since neither the corresponding likelihoods nor the marginal likelihoods are available. However, we can estimate both by adapting the approximate exchange algorithm (AEA) as proposed by \cite{cai:fri11}. 
Algorithm~\ref{alg:AEA} outlines the AEA algorithm for posterior sampling of parameters $\boldsymbol{\zeta}^\mathcal{F}$ and $\boldsymbol{\zeta}^\mathcal{P}$ given observations $\mathbf{z}_{0:T}$ and $\mathbf{x}_{0:T}$. Sampling from $\pi(\boldsymbol{\xi}^\mathcal{F},\boldsymbol{\xi}^\mathcal{P} \mid \mathbf{x}_{0:T})$ follows a similar procedure, but omits the dependence on signed edge states $\mathbf{z}_t$ during proposal generation for interaction structures $\mathbf{x}^{\mathcal{F}'}$ and $\mathbf{x}^{\mathcal{P}'}$.

\begin{algorithm}
\caption{Approximate exchange algorithm for  $\pi(\boldsymbol{\zeta}^\mathcal{F},\boldsymbol{\zeta}^\mathcal{P} \mid \mathbf{z}_{0:T}, \mathbf{x}_{0:T})$}\label{alg:AEA}
\begin{algorithmic}[1]
\Require Initial values $\boldsymbol{\zeta}^{\mathcal{F}(0)}, \boldsymbol{\zeta}^{\mathcal{P}(0)}$
\Ensure Samples $\boldsymbol{\zeta}^{\mathcal{F}(i)}, \boldsymbol{\zeta}^{\mathcal{P}(i)}$ for $i = 1, \dots, I$ iterations
\State Compute $\mathbf{x}_t^\mathcal{F}$ and $\mathbf{x}_t^\mathcal{P}$ for all $t = 1, \dots, T$
\For{$i = 1$ to $I$}
    \State Propose $\boldsymbol{\zeta}^{\mathcal{F}'} \sim h_\mathcal{F}(\cdot),$ and $\boldsymbol{\zeta}^{\mathcal{P}'} \sim h_\mathcal{P}(\cdot)$
    \For{$t = 1$ to $T$}
        \State Simulate $\mathbf{z}_t^{\mathcal{F}'} \sim p\left(\cdot \mid \mathbf{x}_t^\mathcal{F}, \mathbf{y}_{t-1}, \boldsymbol{\zeta}^{\mathcal{F}'} \right)$
        \State Simulate $\mathbf{z}_t^{\mathcal{P}'} \sim p\left(\cdot \mid \mathbf{x}_t^\mathcal{P}, \mathbf{y}_{t-1}, \boldsymbol{\zeta}^{\mathcal{P}'} \right)$
        \State Set $\mathbf{z}_t' = \mathbf{z}_t^{\mathcal{P}'} \cup \left(\mathbf{z}_t^{\mathcal{F}'} \backslash \mathbf{z}_t\right)$
    \EndFor
    \State Compute acceptance probability:
    \Statex \quad \begin{equation}\label{eq:ar}
    \alpha = \min \left(1, 
        \prod_{t=1}^T 
        \frac{\tilde{f}\left(\mathbf{z}_t \mid 
            \mathbf{x}_t, \mathbf{y}_{t-1}, 
            \boldsymbol{\zeta}^{\mathcal{F}'}, 
            \boldsymbol{\zeta}^{\mathcal{P}'}\right)}
            {\tilde{f}\left(\mathbf{z}_t \mid 
            \mathbf{x}_t, \mathbf{y}_{t-1}, 
            \boldsymbol{\zeta}^{\mathcal{F}(i-1)},
            \boldsymbol{\zeta}^{\mathcal{P}(i-1)}\right)}
        \times
        \frac{\pi\left(\boldsymbol{\zeta}^{\mathcal{F}'}, \boldsymbol{\zeta}^{\mathcal{P}'}\right)}
             {\pi\left(\boldsymbol{\zeta}^{\mathcal{F}(i-1)}, \boldsymbol{\zeta}^{\mathcal{P}(i-1)}\right)}
        \right)\end{equation}
    \State With probability $\alpha$, set $(\boldsymbol{\zeta}^{\mathcal{F}(i)}, \boldsymbol{\zeta}^{\mathcal{P}(i)}) \gets (\boldsymbol{\zeta}^{\mathcal{F}'}, \boldsymbol{\zeta}^{\mathcal{P}'})$
    \State Otherwise, set $(\boldsymbol{\zeta}^{\mathcal{F}(i)}, \boldsymbol{\zeta}^{\mathcal{P}(i)}) \gets (\boldsymbol{\zeta}^{\mathcal{F}(i-1)}, \boldsymbol{\zeta}^{\mathcal{P}(i-1)})$
\EndFor
\end{algorithmic}
\end{algorithm}

To address the sampling challenges arising from highly correlated parameters, including correlations between the formation and persistence parameter groups, we employed an adaptive Metropolis–Hastings (AMH) proposal scheme \citep{haa01}. Although the adaptive direction sampling scheme (used by default in \citealt{cai:fri11}) necessitates multiple chains and is less efficient over the long run, it remains valuable for sampling initial parameter values during the burn-in period.

The approximate likelihood ratio in~\eqref{eq:ar} is computed as follows:
$$
\begin{aligned}
\prod_{t=1}^T &
	   \frac{\tilde{f}\left(\mathbf{z}_t \mid 
	         	\mathbf{x}_t, \mathbf{y}_{t-1}, 
	         	\boldsymbol{\zeta}^{\mathcal{F}'}, 
	         	\boldsymbol{\zeta}^{\mathcal{P}'}\right)}
	        {\tilde{f}\left(\mathbf{z}_t \mid 
	            \mathbf{x}_t, \mathbf{y}_{t-1}, 
	            \boldsymbol{\zeta}^{\mathcal{F}(i-1)},
	            \boldsymbol{\zeta}^{\mathcal{P}(i-1)}
	              \right)}
\\
&\hspace{0.38cm} = 
\prod_{t=1}^T
	   \frac{p\left(\mathbf{z}_t \mid 
	         	\mathbf{x}_t, \mathbf{y}_{t-1}, 
	         	\boldsymbol{\zeta}^{\mathcal{F}'}, 
	         	\boldsymbol{\zeta}^{\mathcal{P}'}\right)}
	        {p\left(\mathbf{z}_t \mid 
	            \mathbf{x}_t, \mathbf{y}_{t-1}, 
	            \boldsymbol{\zeta}^{\mathcal{F}(i-1)},
	            \boldsymbol{\zeta}^{\mathcal{P}(i-1)}\right)} 
	   \times
	   \frac{p\left(\mathbf{z}'_t \mid 
	            \mathbf{x}_t, \mathbf{y}_{t-1}, 
	            \boldsymbol{\zeta}^{\mathcal{F}(i-1)},
	            \boldsymbol{\zeta}^{\mathcal{P}(i-1)}\right)}
	        {p\left(\mathbf{z}'_t \mid 
	         	\mathbf{x}_t, \mathbf{y}_{t-1}, 
	         	\boldsymbol{\zeta}^{\mathcal{F}'}, 
	         	\boldsymbol{\zeta}^{\mathcal{P}'}\right)}
\\
&\hspace{0.38cm} = 
\exp 
    \left\{ \left(\boldsymbol{\zeta}^{\mathcal{F}(i-1)}-\boldsymbol{\zeta}^{\mathcal{F}'}\right)^\top 
    \sum_{t=1}^T \left[
    \mathbf{s}\left(\mathbf{z}^{\mathcal{F}'}_t; \mathbf{x}^{\mathcal{F}}_t, \mathbf{y}_{t-1} \right) - 
    \mathbf{s}\left(\mathbf{z}^{\mathcal{F}}_t; \mathbf{x}^{\mathcal{F}}_t, \mathbf{y}_{t-1} \right)
    \right] \right.\\
&\hspace{1cm}\quad\left. 
  + \left(\boldsymbol{\zeta}^{\mathcal{P}(i-1)}-\boldsymbol{\zeta}^{\mathcal{P}'}\right)^\top 
    \sum_{t=1}^T \left[
    \mathbf{s}\left(\mathbf{z}^{\mathcal{P}'}_t; \mathbf{x}^{\mathcal{P}}_t, \mathbf{y}_{t-1} \right) -
    \mathbf{s}\left(\mathbf{z}^{\mathcal{P}}_t; \mathbf{x}^{\mathcal{P}}_t, \mathbf{y}_{t-1}\right)
    \right] \right\}.
\end{aligned}$$

A key advantage of this approach is that network sampling from the likelihood can be conducted with significantly fewer auxiliary iterations and lower computational cost, thanks to the sampling space constraint and reduced complexity of the statistics which are binary.

\subsection{Software} The \textbf{B2Lstergm} package for R \citep{R}, developed alongside this paper, represents a significant extension of existing tools for Bayesian inference on temporal network models. It builds on and extends the modelling and computational capabilities of the \textbf{Bergm} package \citep{cai:bou:kra:fri22, Bergm:package}, which implements Bayesian inference for static ERGMs. Additionally, the \textbf{multi.ergm} package \citep{multi.ergm}, which supports multi-layer ERGMs \citep{kri:koe:mar20}, is employed here to simulate network statistics corresponding to the interaction and conditional sign processes of the two separable likelihood components described in Equation~\ref{eq:sep_stergm_2}. This simulation relies on the efficient MCMC algorithms implemented in the \textbf{ergm} package \citep{ergm, ergm4, ergm:package}. 
Notably, although the \texttt{tergm()} function within the \textbf{tergm} package \citep{tergm:package} typically relies on stochastic maximum likelihood estimation (MLE), it can alternatively use contrastive divergence when computational efficiency is a priority or convergence issues arise. 
The current implementation of the \textbf{B2Lstergm} package can handle networks with a few hundred interacting edges. For larger networks, however, more efficient approximate methods, such as the pseudo-posterior adjustment proposed in \cite{bou:fri:mai17}, would be necessary.

\section{Comparative simulation study}\label{sim}

Among existing approaches, only the TERGM by \citet{fri:meh:thu:kau22} and our approach directly extend the ERGM framework to signed networks. In this simulation study, we compare three modelling approaches with the aim of examining how their interpretative properties differ. 

We consider the TERGM proposed by \citet{fri:meh:thu:kau22}, described in Section~\ref{sec:multi_ergms}, as well as our 2-layer STERGM. In addition, we introduce an intermediate modelling approach: a separable temporal extension of the TERGM in \citet{fri:meh:thu:kau22}. In this STERGM formulation, the transition probability is defined as $$p\left(\mathbf{y}_t \mid \mathbf{y}_{t - 1}, \boldsymbol{\vartheta}^\mathcal{F}, \boldsymbol{\vartheta}^\mathcal{P} \right) = p\left(\mathbf{y}^\mathcal{F}_t \mid  \mathbf{y}_{t - 1}, \boldsymbol{\vartheta}^\mathcal{F} \right) \times p\left(\mathbf{y}^\mathcal{P}_t \mid \mathbf{y}_{t - 1}, \boldsymbol{\vartheta}^\mathcal{P}  \right).$$
It is important to note that when $\mathbf{x}$ is complete, i.e., all nodes interact with all others, the STERGM and the 2-layer STERGM coincide, as $\mathbf{y}$ is binary, encoding the sign of each edge, and therefore equivalent to $\mathbf{z}.$

To better illustrate the interpretative differences between the three approaches, we focus on a simple specification including density and positive transitivity terms. Specifically, we generate a random graph on $40$ nodes for the initial interaction network $\mathbf{x}_1$, where each interaction occurs with probability $0.2$, and each realised edge is independently assigned a positive or negative sign with probability $0.5$. The interaction network $\mathbf{x}_2$ is then generated using a TERGM, with both the baseline edge density and the change-of-state density parameters set to $-2$. Conditional on the transition from $\mathbf{x}_1$ to $\mathbf{x}_2$ we simulate a 2-layer STERGM to generate $\mathbf{z}_2$ specified as follows:
$$
{\zeta_{\texttt{edges}^+}^\mathcal{F}} = 0, \qquad
{\zeta_{\texttt{gwesf}^+}^\mathcal{F}}(\alpha) = 0.2, \qquad
{\zeta_{\texttt{edges}^+}^\mathcal{P}} = 0, \qquad
{\zeta_{\texttt{gwesf}^+}^\mathcal{P}}(\alpha) = 0.
$$
Here, $\texttt{edges}^+$ denotes the number of positive edges in the formation and dissolution processes, and $\texttt{gwesf}^+$ is the geometrically weighted positive edgewise shared friends (see Figure~\ref{SBT_stats}) representing the statistics related to positive triadic closure. We set the decay parameter $\alpha$ for the $\texttt{gwesf}^+$ terms to $0.6$ In practice, the only non-zero effect is therefore the formation parameter ${\zeta_{\texttt{gwesf}^+}^\mathcal{F}}$, which encourages the creation of positive triangles from $t_1$ to $t_2$.

The TERGM approach includes terms for the stability of positive edges, changes in positive edges, and changes in transitive triads. The STERGM version of \citet{fri:meh:thu:kau22} uses the same type of statistics employed in the simulation, as does the 2-layer STERGM.

\begin{figure}[htp]
\centering
\includegraphics[scale=0.4]{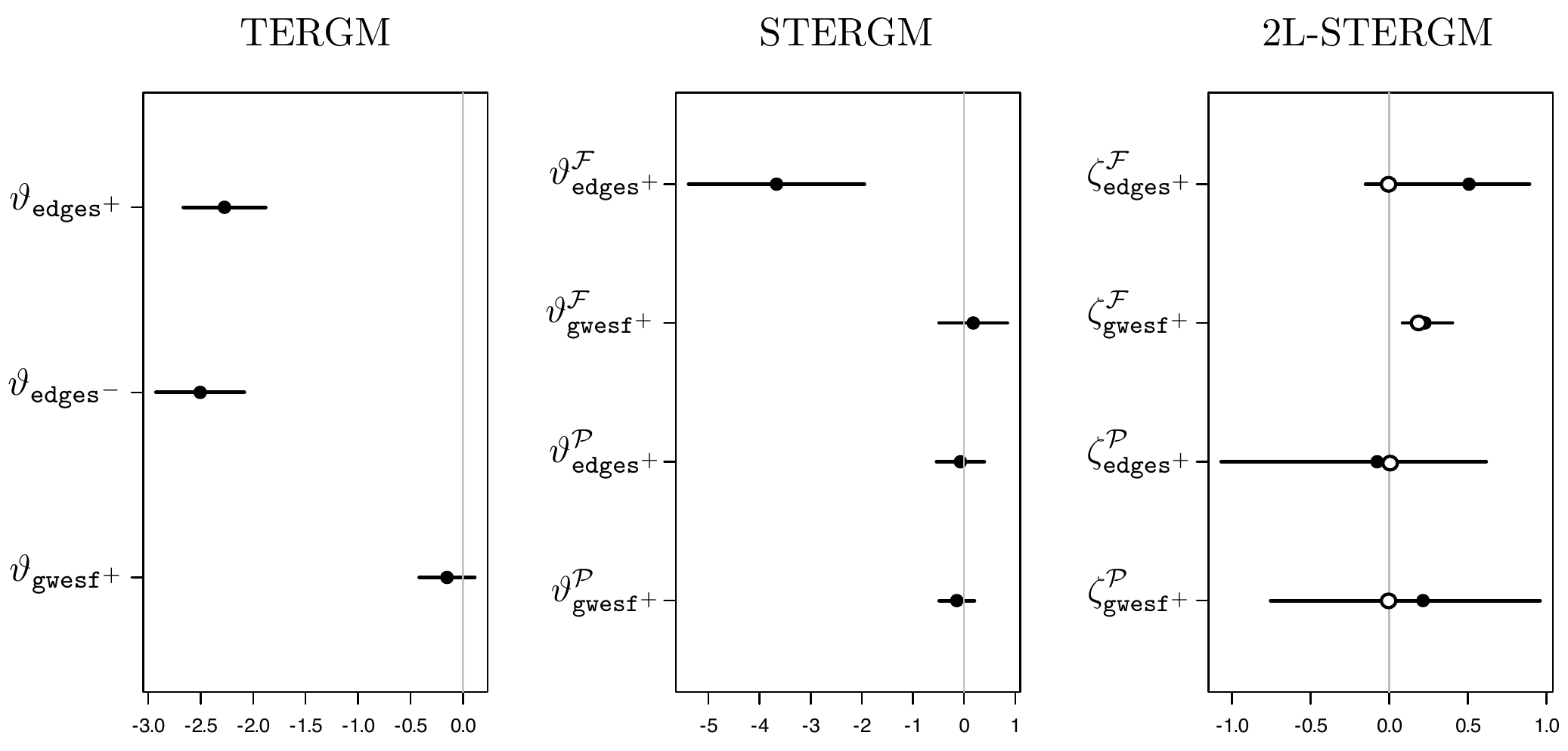}
\caption{MLE estimates (black dots) with 95\% confidence intervals are shown for the TERGM and STERGM based on \citet{fri:meh:thu:kau22}, while the 2-layer STERGM displays posterior means (black dots) with 95\% credible intervals. In the 2-layer STERGM plot, the white dots with black outline indicate the true parameter values used to generate the data.}
\label{fig:sim_comp}
\end{figure}

In Figure~\ref{fig:sim_comp}, we can observe the MLE estimates and their corresponding 95\% confidence intervals for both the TERGM and STERGM approaches, computed using the \textbf{ergm.sign} package for R \citep{ergm.sign}. We also show the posterior means and 95\% credible intervals for the two-layer STERGM, obtained with the \textbf{B2Lstergm} package using flat priors on all parameters, in order to make the comparison easier.

We can see that the positive and negative density parameters in the TERGM correctly capture the drop in overall density caused by the reduced level of interaction from $t_1$ to $t_2$. The parameter $\vartheta_{\texttt{gwesf}^+}$, associated with positive triangles, is not significantly different from zero. In this model, without additional controls for other forms of transitive structure, it is not possible to determine whether this value reflects a real counteracting effect, or whether the change instead arises from substantially greater formation or proportionally less dissolution of transitive structures, relative to the others. In practice, this model determines dyad change rates according to the observed probability of each dyad switching state. Therefore, when the initial network has a density below $0.5$, there are more opportunities for formation, whereas a density above $0.5$ results in more opportunities for dissolution.

In the STERGM, the parameter ${\vartheta_{\texttt{edges}^+}^\mathcal{F}}$ is negative, as it reflects the fact that the proportion of positive edges formed between $t_1$ and $t_2$ is negative due to the non significant increase in the proportion of positive edges in the formation process (relative to the proportion of negative edges and empty dyads). The parameter ${\vartheta_{\texttt{gwesf}^+}^\mathcal{F}}$ is not significant.
If we add a control for transitive interaction in the formation process, ${\vartheta_{\texttt{gwesf}^+}^\mathcal{F}}$ becomes significantly positive ($5.46$ with a standard error of about $0.17$), balancing a significantly negative parameter ($-5.65$ with a standard error of about $0.16$). We can therefore conclude that this categorical model captures the increase in positive transitive structures, but we cannot determine precisely whether this is due to the formation of genuinely new positive interactions or to the transformation of existing non-positive configurations into positive ones. In other words, we cannot clearly separate the effect of sign changes from the formation interaction process itself.

The 2L-STERGM model, through the parameter ${\zeta_{\texttt{gwesf}^+}^\mathcal{F}}$, captures the increase in the proportion of positive transitive relations at the expense of other types of triadic signed configurations, allowing us to deduce that there has been a rise in positive transitive structures, net of overall structural changes in the interaction network. The greater uncertainty in the persistence process, arises from the reduced sample space determined by the low number of persistent interactions from $t_1$ to $t_2$.

\section{Application to US Congressional data}\label{app}

We employ our model to examine the evolution of political cooperation and opposition in the United States Congress using the signed network data inferred by \cite{zac14,zac20} from bill co-sponsorship data via stochastic degree sequence model (SDSM). 

Our focus is on the 99th-101st Congresses (1985-1991), characterised by a hung government with a split majority. This period saw a republican president, Ronal Reagan, governing alongside a Democratic majority in the Senate. An examination of the Senate behaviour during this time can provide valuable insights into the dynamics of the Senate and its role in shaping policy during a period of significant upheaval \citep{rohde1991}. This period encompasses several pivotal events, including the collapse of the Soviet Union (1989-1991), the Gulf War (1990-1991), and the Savings and Loan Crisis (1989-1991).

When Congressional co-sponsorship data are modelled using the SDSM framework, each legislator has expected positive and negative degree tendencies. For each dyad, the probability of a positive, negative, or no edge is computed based on the product of their respective tendencies. Edges are then assigned probabilistically: dyads with higher positive tendencies are more likely to become positive edges, those with higher negative tendencies are more likely to become negative edges, and others remain unconnected. This produces a signed network that reflects legislators' expected co-sponsorship and avoidance patterns.

We concentrate our analysis on the 70 Senators who remained in office throughout the 99th-101st Congresses, in order to examine the behaviour of the network effects related to their party affiliation and structural balance.

\begin{figure}[htp]
\centering
\includegraphics[scale=0.5]{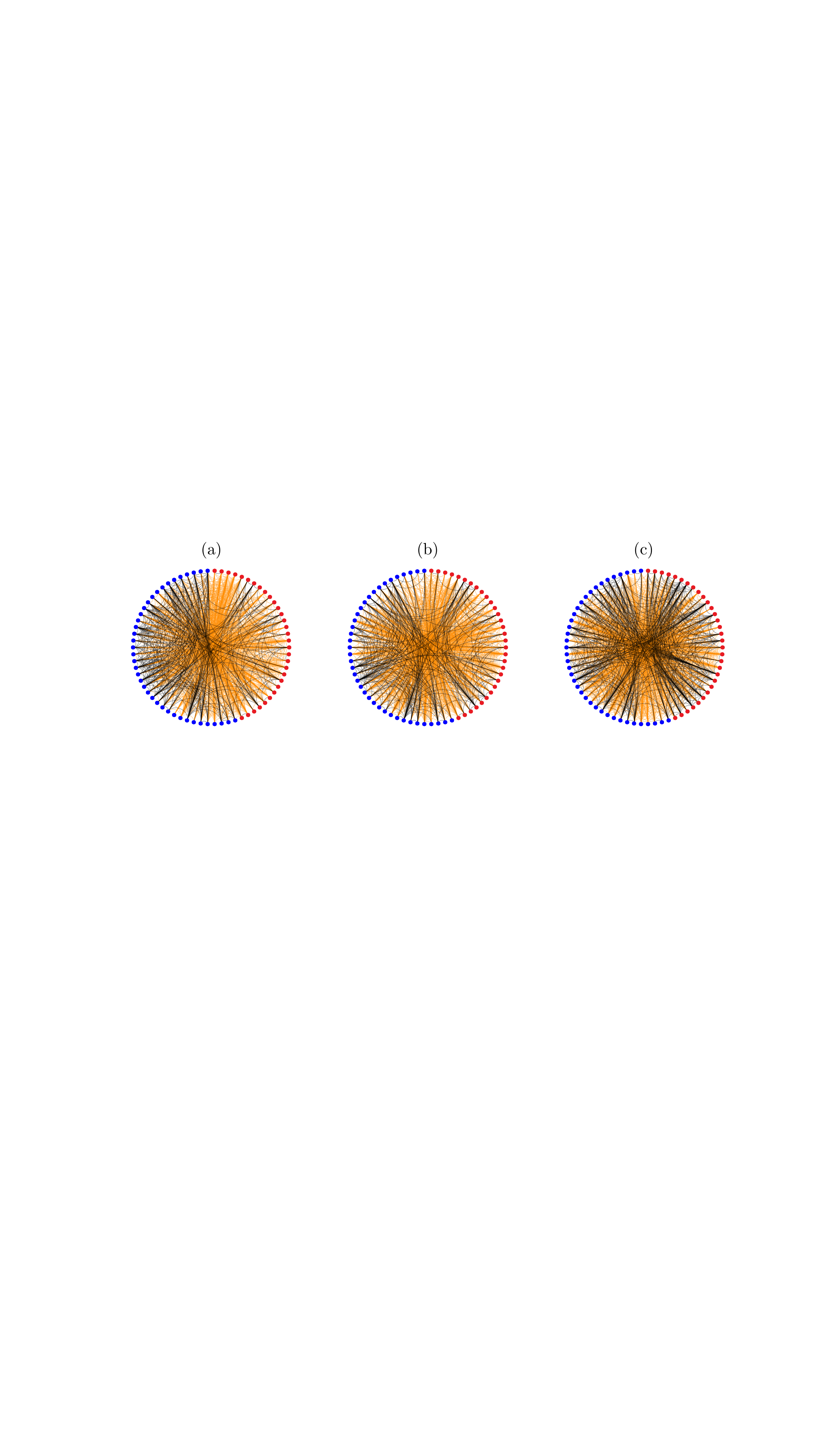}\\[-0.5cm]
\caption{US Senators signed network. Blue and red nodes represent Democratic and Republican Senators, respectively. Black and orange edges represent positive and negative relationships, respectively. Graphs: (a) signed relationships during the 99th Congress (initial state); (b) signed relationships during the 100th Congress; (c) signed relationships during the 101st Congress.}
\label{fig:senators}
\end{figure}

The observed signed networks displayed in Figure~\ref{fig:senators} exhibit considerable interaction density. Specifically, the formation interaction density rises from just below 0.60 to nearly 0.61, while the persistence interaction density increases from 0.26 to almost 0.31. Most interactions are negative, even within parties, across both the formation and persistence processes.

\subsection{Model specification}\label{app:specs}

We adopt a time-homogeneous 2-layer STERGM where the formation and persistence configurations are the same and the list of statistics used is specified below. Obviously, in the case of excessive non-stationarity across multiple network observations, the use of a homogeneous model could prove problematic.

As mentioned in the previous section, our focus is to assess the extent of the importance of weak balance structures, and thus we include (non-degenerate) configurations corresponding to the weakly balanced triads (Figure~\ref{fig:SBT}) in both the formation and persistence processes.

The interaction network exhibits such high density that it lacks discernible structural heterogeneity. Consequently, we restrict our analysis to the conditional signed process and the posterior distribution of its associated parameters $\pi(\boldsymbol{\zeta}^\mathcal{F},\boldsymbol{\zeta}^\mathcal{P} \mid \mathbf{z}_{0:T}, \mathbf{x}_{0:T})$. We define a 2-layer STERGM using the following network statistics:
\begin{itemize}
\item[1.] number of positive edges, \texttt{edges}$^+,$ measuring the density of positive interactions;
\item[2.] number of positive edges within the Republican party, \texttt{homophily}$^+(\texttt{rep}),$ measuring the density of positive interactions within the Republican party;
\item[3.] geometrically-weighted positive degrees, \texttt{gwdegree}$^+(\alpha),$ measuring the tendency of actors to form multiple positive edges;
\item[4.] geometrically-weighted positive edgewise shared friends, \texttt{gwesf}$^{+}(\alpha),$ capturing the positive triadic closure configurations;
\item[5.] geometrically-weighted positive edgewise shared enemies, \texttt{gwese}$^{+}(\alpha),$ capturing the tendency for negative triadic closure among positive edges;
\item[6.] geometrically-weighted negative edgewise shared enemies, \texttt{gwese}$^{-}(\alpha),$ capturing negative triadic closure configurations.
\end{itemize}

We set the decay parameter $\alpha$ to $0.2$ for the geometrically-weighted degree statistic, and to $0.6$ for all geometrically-weighted edgewise shared partner-based statistics \citep{sni:pat:rob:han06}.

We assign independent, weakly informative Normal priors to all model parameters, incorporating mild guidance from the structural characteristics of the initial network. Specifically, the prior means for both $\zeta^{\mathcal{F}}_{\text{edges}^+}$ and $\zeta^{\mathcal{P}}_{\text{edges}^+}$ are set to $–1$, reflecting a prior belief that the baseline probability of a positive edge (not involved in triadic or degree-based configurations) is low, but without imposing a strong constraint on its magnitude. Similarly, the prior mean for $\zeta^{\mathcal{F}}_{\text{gwesf}^+}(\alpha)$ is set to reflect moderate triadic closure, consistent with expectations of structural balance in cooperative interactions. All priors are assigned a large variance of 25, ensuring they remain weakly informative, vague enough to let the data dominate inference, while still preventing numerical instabilities and implausible parameter values during MCMC estimation.

\subsection{Posterior inference}\label{results}

\begin{figure}[htp]
\centering
\includegraphics[scale=0.7]{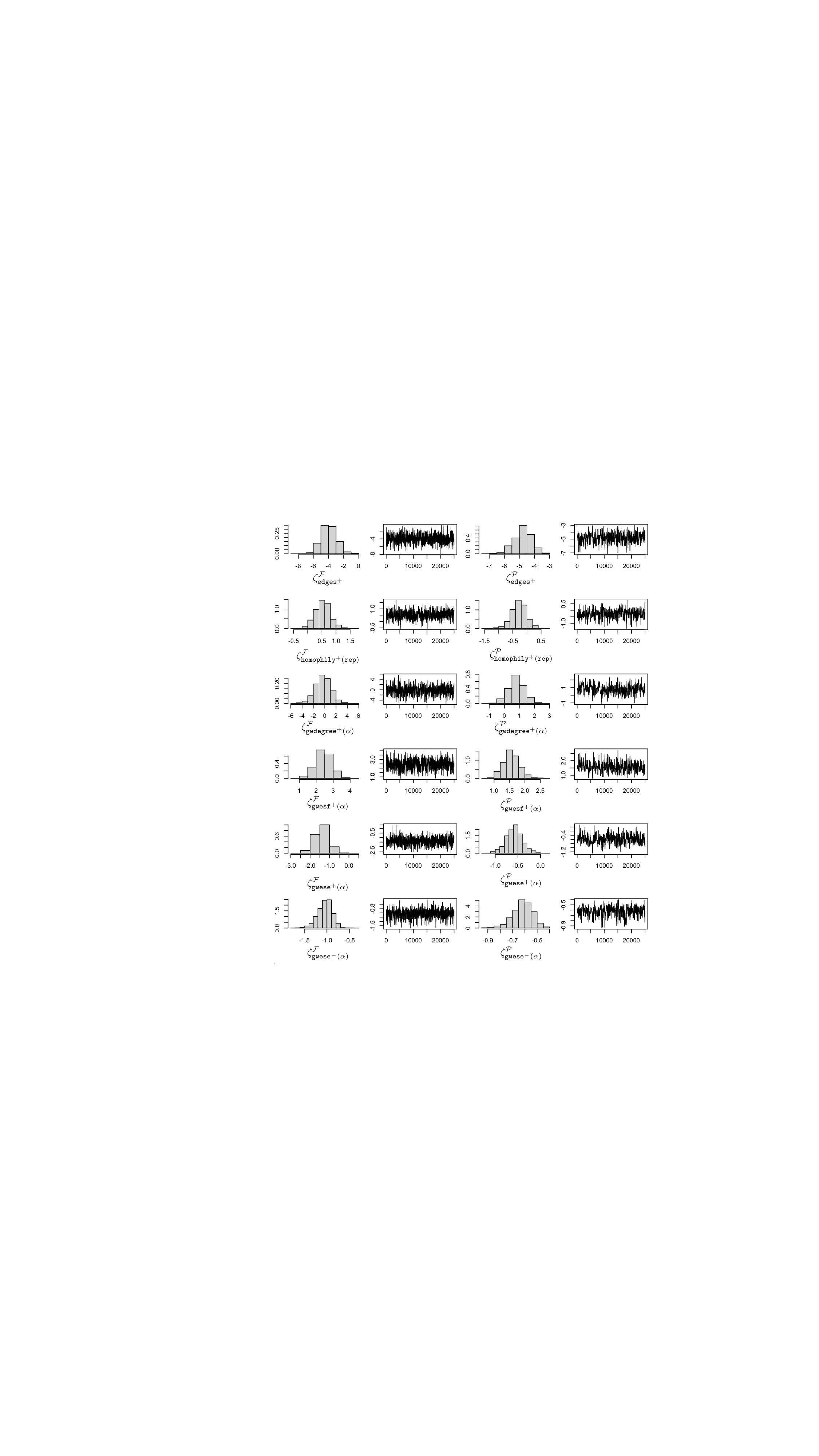}\\[-0.5cm]
\caption{MCMC histograms and trace plots for each marginal posterior parameter distribution.}
\label{fig:mcmc_traces}
\end{figure}

Posterior inference is conducted using the adaptive AEA, as outlined in Algorithm~\eqref{alg:AEA}. Each iteration employs $5000$ auxiliary draws for the network simulation, and the main Markov chain comprises $35000$ iterations, with the initial $10000$ discarded as burn-in.  MCMC trace plots (Figure~\ref{fig:mcmc_traces}) of the simulated posterior parameters indicate that the MCMC chain converges to the stationary posterior distribution. Table~\ref{tab:post} presents a summary of the posterior density parameter estimates.

\begin{figure}[htp]
\centering
\includegraphics[scale=0.5]{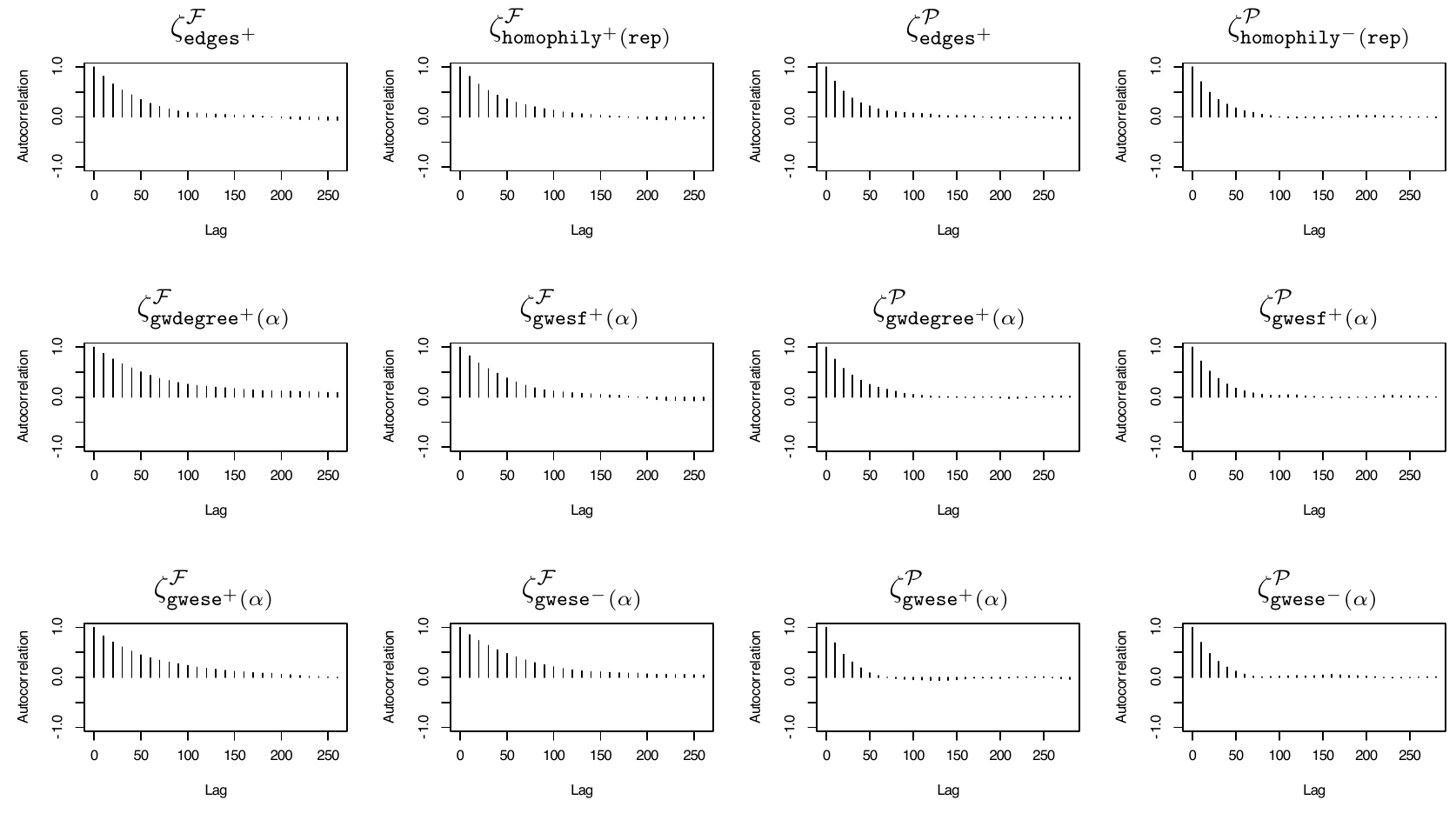}\\
\caption{MCMC autocorrelation plots for each parameter.}
\label{fig:mcmc_rho}
\end{figure}

The quality of the posterior estimates was evaluated using autocorrelation plots (Figure~\ref{fig:mcmc_rho}) and effective sample sizes. All parameters exhibited effective sample sizes greater than 200, suggesting adequate chain convergence and mixing.

\begin{figure}[htp]
\centering
\includegraphics[scale=0.3]{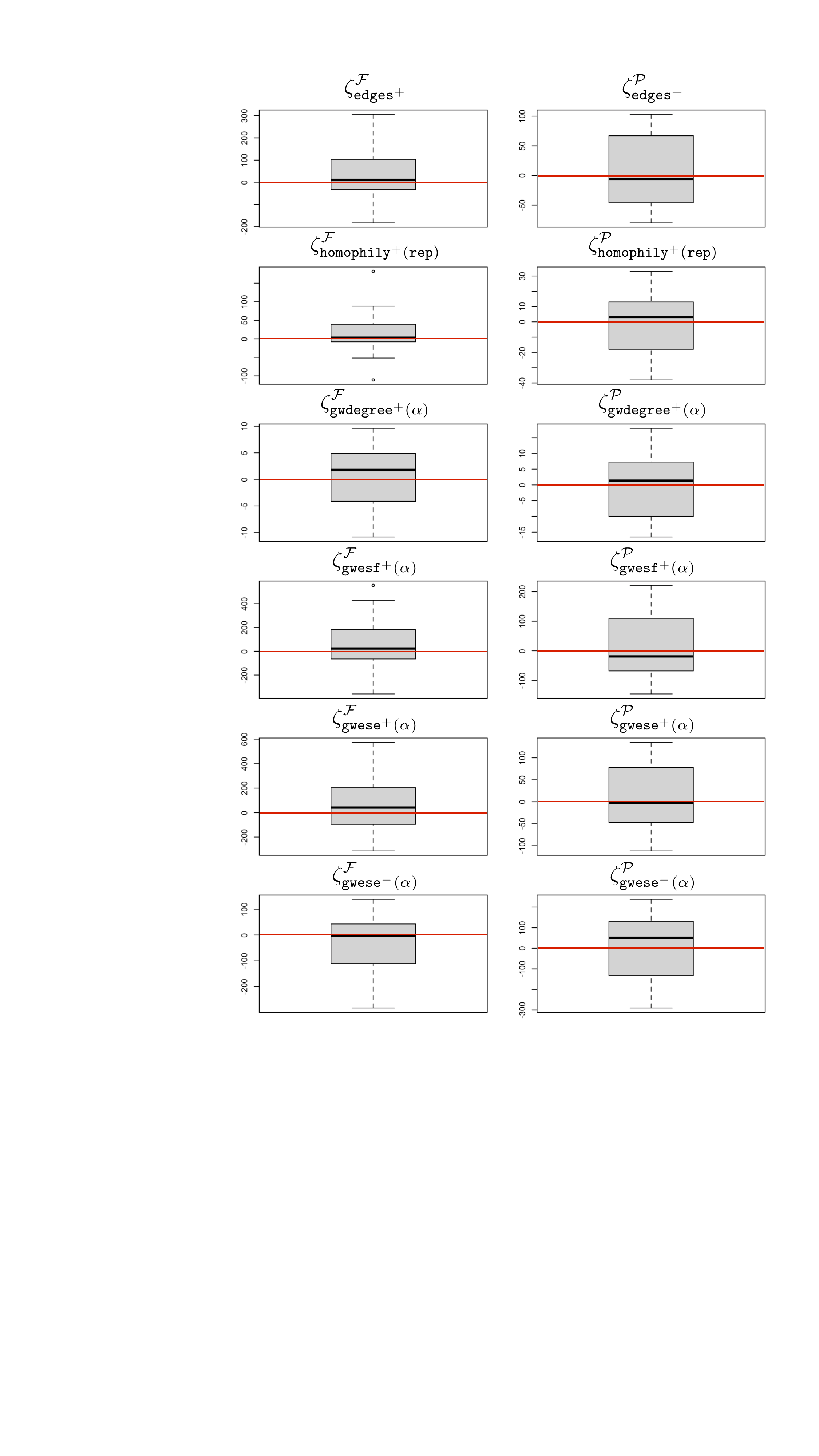}\\[-0.5cm]
\caption{Posterior predictive boxplots of centred cumulative differences over time between simulated and observed network statistics: $\sum_{t = 1}^T \mathbf{s}_q\left(\mathbf{z}^{\mathcal{U}'}_t; \mathbf{x}^{\mathcal{U}}_t, \mathbf{y}_{t-1}, \zeta_q^{\mathcal{U}} \right) - \mathbf{s}\left(\mathbf{z}^{\mathcal{U}}_{t}; \mathbf{x}^{\mathcal{U}}_t, \mathbf{y}_{t-1}, \zeta_k^{\mathcal{U}} \right), \mathcal{U} = \{{\mathcal{F}, \mathcal{P}}\}, q = 1, \cdots, Q.$}
\label{fig:post_pred}
\end{figure}

Posterior predictive checks are performed to assess how well the model captures the observed data. Simulated network statistics were generated from the posterior predictive distribution and compared to the observed statistics using boxplots (Figure~\ref{fig:post_pred}). In these plots, the medians of the posterior predictive statistics are close to $0,$ indicating that the observed data are consistent with what the model predicts. This suggests that the model provides a good fit to the data in terms of model statistics.

\begin{table}[h]
\caption{Summary of the posterior parameter density estimates of the conditional sign process.}\label{tab:post}\setlength{\tabcolsep}{7.5pt}
\begin{tabular}{l|cccc|cccc}
\hline
 & \multicolumn{4}{|c}{Formation ($\mathcal{F}$)} & \multicolumn{4}{|c}{Persistence ($\mathcal{P}$)} \\
\cline{2-9}  
\multicolumn{1}{c|}{Parameters}  &  Mean & 2.5\% & Median & 97.5\% & Mean & 2.5\% & Median & 97.5\% \\
\cline{1-9} 
$\zeta_{\texttt{edges}^{+}}$                    & -3.95 & -5.81 & -3.98 & -2.00 & -4.77 & -5.90 & -4.76 & -3.68 \\ 
$\zeta_{\texttt{homophily}^{+}(\texttt{rep})}$  &  0.53 &  0.02 &  0.53 &  1.06 & -0.28 & -0.83 & -0.27 &  0.25 \\ 
$\zeta_{\texttt{gwdegree}^{+}(\alpha = 0.2)}$   & -0.29 & -3.11 & -0.31 &  2.49 &  0.81 & -0.28 &  0.79 &  1.87 \\ 
$\zeta_{\texttt{gwesf}^{+}(\alpha = 0.6)}$      &  2.42 &  1.46 &  2.44 &  3.38 &  1.56 &  1.05 &  1.56 &  2.11 \\ 
$\zeta_{\texttt{gwese}^{+}(\alpha= 0.6)}$       & -1.41 & -2.11 & -1.40 & -0.70 & -0.58 & -0.96 & -0.58 & -0.20 \\
$\zeta_{\texttt{gwese}^{-}(\alpha= 0.6)}$       & -1.05 & -1.41 & -1.03 & -0.75 & -0.62 & -0.81 & -0.61 & -0.48 \\
\cline{1-9}                
\end{tabular}
\end{table}

Although the analysis is restricted to the subset of U.S. Senators who served continuously from the 99th to the 101st Congress, the posterior parameter estimates, presented in Table~\ref{tab:post}, yield several key insights into the dynamics of political interactions:

\begin{itemize}

\item The strongly negative 95\% posterior credible intervals for $\zeta_{\texttt{edges}^{+}}$ in both the formation and persistence processes indicate that the baseline probability of forming or maintaining a positive edge is very low. For example, the posterior mean of $\zeta_{\texttt{edges}^{+}} = -3.95$ in the formation process corresponds to a conditional probability of forming a positive edge given an interaction of $\operatorname{logit}^{-1}(-3.95) \approx 0.019$, and for persistence, $\zeta_{\texttt{edges}^{+}} = -4.77$ gives $\operatorname{logit}^{-1}(-4.77) \approx 0.008.$

\item The 95\% posterior credible interval for $\zeta_{\texttt{homophily}^{+}(\texttt{rep})}$ is slightly positive for formation but not significantly different from 0 for persistence. Marginally, this translates to a modest increase in the probability of forming a positive edge among republican senators, e.g., $\operatorname{logit}^{-1}(-3.95 + 0.53) \approx 0.027$, while the effect on persistence remains negligible.

\item The estimates for $\zeta_{\texttt{gwdegree}^{+}(\alpha = 0.2)}$ are not significantly different from zero in either process, implying that high-degree nodes do not systematically form or maintain more positive edges. This suggests a relatively even distribution of interactions across the network.

\item The consistently positive 95\% posterior credible intervals for $\zeta_{\texttt{gwesf}^{+}(\alpha = 0.6)}$ support the role of triadic closure in both forming and sustaining positive interactions. For instance, adding one unit of balanced triadic closure increases the conditional probability of positive formation from $\approx 0.019$ to $\operatorname{logit}^{-1}(-3.95 + 2.42) \approx 0.18$ and the probability of persistence from $\approx 0.008$ to $\operatorname{logit}^{-1}(-4.77 + 1.56) \approx 0.039$.

\item The negative posterior estimates for $\zeta_{\texttt{gwese}^{+}(\alpha = 0.6)}$ and $\zeta_{\texttt{gwese}^{-}(\alpha = 0.6)}$, where the former corresponds to a balanced triadic configuration and the latter to an unbalanced one, indicate that these structures reduce the likelihood of forming or maintaining positive edges. For example, a dyad involved in an unbalanced negative triad would have its conditional formation probability reduced from $\approx 0.019$ to $\operatorname{logit}^{-1}(-3.95 - 1.05) \approx 0.0067.$ 
\end{itemize}

\subsection{Marginal estimation}

The network interaction structure $\mathbf{x}_{0:T}$ was analysed by estimating the following STERGM posterior
$$
\pi\left(\boldsymbol{\xi}^\mathcal{F}, \boldsymbol{\xi}^\mathcal{P} \mid \mathbf{x}_{0:T}\right)
\propto \prod_{t = 1}^T  \exp \left\{{\boldsymbol{\xi}^\mathcal{F}}^{\top} \mathbf{s}\left(\mathbf{x}_t^\mathcal{F}; \mathbf{y}_{t-1} \right) \right\}\;
\exp \left\{{\boldsymbol{\xi}^\mathcal{P}}^{\top} \mathbf{s}\left(\mathbf{x}_t^\mathcal{P}; \mathbf{y}_{t-1} \right) \right\}\; \pi\left(\boldsymbol{\xi}^\mathcal{F}\right)\; \pi\left(\boldsymbol{\xi}^\mathcal{P}\right)
$$
using a similar version of Algorithm~\ref{alg:AEA}. For both the formation and persistence, we used the following three network statistics that are matching the configurations used in the conditional signed process: (1) $\texttt{edges}$, the baseline propensity for any interaction to form or persist; (2) $\texttt{homophily(rep)},$ the propensity for interactions between republican Senators to form or persist; (3) $\texttt{gwesp}(\alpha = 0.2),$ the tendency for interactions to form or persist in closed triads; and (4) $\texttt{gwdegree}(\alpha = 0.6),$ the influence of node degree on the formation and persistence of edges. Posterior sampling was performed using a prior specification and initialisation procedure similar to that described in Section~\ref{results}. 

\begin{table}[h]
\caption{Summary of the posterior parameter density estimates of the interaction process.}\label{tab:post_x}\setlength{\tabcolsep}{7.5pt}
\begin{tabular}{l|cccc|cccc}
\hline
 & \multicolumn{4}{|c}{Formation ($\mathcal{F}$)} & \multicolumn{4}{|c}{Persistence ($\mathcal{P}$)} \\
\cline{2-9}  
\multicolumn{1}{c|}{Parameters}  &  Mean & 2.5\% & Median & 97.5\% & Mean & 2.5\% & Median & 97.5\% \\
\cline{1-9} 
$\xi_{\texttt{edges}}$                    & -6.40 & -7.45 & -6.38 & -5.46 & -4.78 & -5.46 & -4.80 & -4.12 \\ 
$\xi_{\texttt{homophily}(\texttt{rep})}$  &  0.08 & -0.15 &  0.08 &  0.31 &  0.06 & -0.30 &  0.06 &  0.43 \\ 
$\xi_{\texttt{gwdegree}(\alpha = 0.2)}$   & -1.18 & -3.65 & -1.11 &  0.89 & -0.10 & -0.94 & -0.08 &  0.66 \\ 
$\xi_{\texttt{gwesp}(\alpha = 0.6)}$      &  2.52 &  2.04 &  2.51 &  3.07 &  1.55 &  1.24 &  1.56 &  1.88 \\ 
\cline{1-9}                
\end{tabular}
\end{table}

The results summarised in Table~\ref{tab:post_x} indicate that new interaction formation was relatively infrequent, with a strongly negative baseline tendency ($\xi^{\mathcal{F}}_{\texttt{edges}}$), suggesting that in the absence of other effects, new edges were unlikely to form. There was only weak evidence of Republican homophily in edge formation ($\xi^{\mathcal{F}}_{\texttt{homophily}(\texttt{rep})}$), indicating minimal preference for within-group connections. Actors with higher existing degrees were somewhat less likely to create additional edges ($\xi^{\mathcal{F}}_{\texttt{gwdegree}(\alpha)}$) consistent with degree saturation effects limiting new edge formation. By contrast, there was strong and credible evidence of transitive closure in the formation process ($\xi^{\mathcal{F}}_{\texttt{gwesp}(\alpha)}$), indicating that new edges tend to form within cohesive triads.

In contrast, existing interactions display a strong tendency to persist, although overall edge stability remained selective ($\xi^{\mathcal{P}}_{\texttt{edges}}$). Homophily again played little role in persistence ($\xi^{\mathcal{P}}_{\texttt{homophily}(\texttt{rep})}$). Degree effects were near zero ($\xi^{\mathcal{P}}_{\texttt{gwdegree}(\alpha)}$), implying that high-degree individuals were only marginally less likely to sustain their existing edges. Transitive closure effects were positive ($\xi^{\mathcal{P}}_{\texttt{gwesp}(\alpha)}$), showing that edges embedded in triadic structures were substantially more stable over time.

Overall, these results suggest a relatively stable network characterised by limited new edge creation, strong clustering tendencies within triads, and modest degree-related constraints on both the formation and maintenance of interactions. New connections tend to emerge within existing cohesive subgroups, while established relationships persist preferentially within these clustered structures.

To obtain marginal probabilities for positive edges, we multiply the interaction probabilities by the conditional probabilities according to Equation~\eqref{eq:sepERGM_2}. 

The estimated probabilities suggest that the formation of new interactions in the network is rare. The strongly negative formation parameter $\xi_{\texttt{edges}}$ implies that, in the absence of additional structural effects, the probability of any new edge emerging is extremely low, $\text{logit}^{-1}(-6.40) \approx 0.0017$. When this is combined with the small conditional probability that a newly formed interaction is positive ($\approx 0.019$), the resulting marginal probability of observing a new positive edge becomes effectively negligible. However, this pattern changes markedly once triadic closure is considered. Incorporating the transitivity effect $\xi_{\texttt{gwesp}(\alpha = 0.6)}$ increases the probability of forming an interaction to about $\operatorname{logit}^{-1}(-6.40 + 2.52) \approx 0.02$, and the probability that such an interaction is positive to roughly $\operatorname{logit}^{-1}(-3.95 + 2.42) \approx 0.18$. The resulting marginal probability of forming a positive edge within a triad ($\zeta_{\texttt{gwesf}^{+}}$) rises to around $0.18 \times \text{logit}^{-1}(-6.40 + 2.52) \approx 0.0036$. This clearly indicates that the formation of positive edges is concentrated within cohesive positive triadic structures.

Persistence follows a similar but slightly stronger pattern. Existing interactions are more likely to continue than new ones are to form, with a baseline persistence probability of around $0.0083$. Yet, the probability that a persisting edge remains positive is very close to 0. When triadic effects are present, both persistence and conditional positivity rise to approximately $\text{logit}^{-1}(-4.78+1.55) \approx  0.038$, resulting in a marginal positive persistence of about $0.038 \times \text{logit}^{-1}(-4.77 + 1.56) \approx0.0015$. These findings imply that while the overall network formation structure remains sparse, positive relationships, though rare, are substantially more likely to both form and endure within tightly clustered triads. Taken together, the results portray a network characterised by isolated sets of stable, mutually reinforcing positive interactions. 

While these estimates provide a useful illustration of the patterns in the network, the true uncertainty around these marginal probabilities may be larger, and credible intervals derived from posterior draws provide a more complete picture.

\section{Discussion}\label{discussion}

This paper introduces a novel separable temporal generative framework for modelling the dynamics of signed networks. By distinguishing between interaction effects and conditional sign processes, our approach preserves the flexibility and interpretability of traditional binary exponential random graph models while remaining consistent with the assumptions of structural balance theory (SBT). 

When modelling signed networks, the separable approach used in this paper is particularly advantageous when formation and dissolution processes are plausibly driven by different mechanisms, and when the goal is to rigorously test structural balance configurations while accounting for the potentially strong influence of overall interaction density on the observed signed edges. In such settings, some signed edges may be unobserved or latent, and failing to account for this can bias the results.

To empirically test SBT, we specifically incorporate endogenous effects that capture patterns predicted by the theory. Nonetheless, the framework remains fully compatible with the broader class of binary statistics commonly used in separable temporal ERGM processes, including exogenous and lagged covariates. This separation allows for a deeper understanding of how network relations evolve over time, making it suitable for complex systems in which the interplay of both positive and negative relationships is critical to the system overall dynamics. When the assumptions underlying separability are not met, alternative approaches that relax one of these assumptions can be employed.

We employed a fully probabilistic Bayesian approach to infer model parameters, utilising an adaptive Metropolis-Hastings approximate exchange algorithm to efficiently estimate the doubly intractable posterior distribution. This method enables us to account for the complexity of signed network structures and provides robust uncertainty estimates despite the computational challenges. 

The Bayesian framework also enables assessment of potential violations by comparing posterior predictive performance across alternative models, for instance, by examining whether a model that allows conditional dependence between the processes yields a substantially better fit.

We analysed the political relationships among U.S. Senators during Ronald Reagan's second term (1985–1989), uncovering the evolving dynamics of political alliances and rivalries within the Senate. By applying our framework with a focus on endogenous structural effects grounded in SBT, we identified key patterns of both supportive and antagonistic alliances, shedding light on the structural mechanisms underpinning shifts in political coalitions.

One limitation of this empirical analysis is that the signed relationships in the network are derived from a projection of the underlying bipartite network, based on the similarity of interaction patterns within that same structure \citep{nea:etal24}. As such, the sign attributed to edges reflects inferred structural similarity rather than directly observed relational content. A more direct approach would involve the use of longitudinal relational data in which tie signs carry an explicit substantive meaning, for example, recorded friendships and antagonistic relationships between individuals.

The flexibility and interpretability of our model open up several avenues for future research. One promising direction involves extending our approach to a continuous-time framework, such as those based on longitudinal ERGMs \citep{kos:lom13,kos:cai:lom15}, which could significantly enhance the model applicability to real-time, dynamic network data. This extension may involve the development of data augmentation strategies in which MCMC algorithms alternate between sampling from the conditional posterior distribution of the model parameters and generating plausible interaction paths that connect consecutive network snapshots. Another direction concerns the adaptation of the methodology to dynamic weighted signed networks, where the transition probabilities of the signed process follow a partially separable temporal model, as recently proposed by \cite{kei:etal23}. Crucially, the feasibility of these extensions will depend on the development of more efficient and scalable estimation algorithms. As network data grow in complexity and temporal resolution, computational bottlenecks become increasingly significant. Advancing the underlying inferential algorithm, through improved MCMC schemes, variational approximations \citep{tan:fri20}, will be essential to ensure the practical usability of these models in applied settings.

\section{Acknowledgments}
The authors thank Marc Schalberger and Cornelius Fritz for sharing an early version of the \textbf{ergm.sign} package, which enabled us to complete the comparative simulation study presented in this paper. We thank the anonymous reviewers for their insightful comments and constructive suggestions, which have substantially improved the clarity and quality of the manuscript.

\section{Data availability}
The dataset and code used in this paper are available for download at
\href{https://github.com/acaimo/B2Lstergm}{github.com/acaimo/B2Lstergm}.

\bibliographystyle{abbrvnat}

\end{document}